\documentclass[twocolumn]{aastex63}
\usepackage{hyperref}
\usepackage{color}
\usepackage{comment}
\usepackage{ifthen}
\usepackage{grffile}
\usepackage[utf8]{inputenc}
\usepackage{bm}
\usepackage{amsmath,mathtools}

\newcommand{\mat}[1]{\mathbf{#1}}
\newcommand{\T}{\mathrm{T}}
\newcommand{\minus}{\scalebox{0.75}[1.0]{$-$}}
\newcommand{\ex}[2][]{\mathbb{E}_{#1}\left[#2\right]}
\newcommand{\var}[2][]{\mathbb{V}_{#1}\left[#2\right]}
\newcommand{\cov}[3][]{\mathbb{C}_{#1}\left[#2,\, #3\right]}
\specialcomment{notes}{\begingroup\sffamily\color{red}}{\endgroup}
\includecomment{comment}

\begin{document}

\defcitealias{Paine:2022fo}{P22}
\defcitealias{Darling:2023ao}{D23}

\title{The HST-Gaia Near-Infrared Astrometric Reference Frame near the Milky Way Galactic Center}

\author[0000-0003-2874-1196]{Matthew W. Hosek Jr.}
\correspondingauthor{Matthew W. Hosek Jr.}
\altaffiliation{Brinson Prize Fellow}
\affiliation{UCLA Department of Physics and Astronomy, Los Angeles, CA 90095}
\email{mwhosek@astro.ucla.edu}

\author[0000-0001-9554-6062]{Tuan Do}
\affiliation{UCLA Department of Physics and Astronomy, Los Angeles, CA 90095}

\author[0000-0002-7476-2521]{Gregory D. Martinez}
\affiliation{UCLA Department of Physics and Astronomy, Los Angeles, CA 90095}

\author[0000-0001-7003-0588]{Rebecca Lewis-Merrill}
\affiliation{UCLA Department of Physics and Astronomy, Los Angeles, CA 90095}

\author[0000-0003-3230-5055]{Andrea M. Ghez}
\affiliation{UCLA Department of Physics and Astronomy, Los Angeles, CA 90095}

\author[0000-0001-9611-0009]{Jessica R. Lu}
\affiliation{Department of Astronomy, 501 Campbell Hall, University of California, Berkeley, CA, 94720}

\author[0000-0001-5972-663X]{Shoko Sakai}
\affiliation{UCLA Department of Physics and Astronomy, Los Angeles, CA 90095}

\author[0000-0003-2861-3995]{Jay Anderson}
\affiliation{Space Telescope Science Institute, 3700 San Martin Drive, Baltimore, MD 21218, USA}

\begin{abstract}
We present the first high-precision proper motion catalog, tied to the International Celestial Reference System (ICRS),
of infrared astrometric reference stars within R $\leq$ 25'' (1 pc) of the central supermassive black hole at the Galactic center (GC).
This catalog contains $\sim$2,900 sources in a highly extinguished region that is inaccessible via Gaia. New astrometric measurements
are extracted from HST observations (14 epochs, 2010 - 2023) and transformed into the ICRS using 40 stars in common with Gaia-DR3.
We implement a new method for modeling proper motions via Gaussian Processes that accounts for systematic errors, greatly improving measurement accuracy.
Proper motion and position measurements reach precisions of $\sim$0.03 mas yr$^{-1}$ and $\sim$0.11 mas, respectively, representing a factor of $\sim$20x
improvement over previous ICRS proper motion catalogs in the region. These measurements define a novel HST-Gaia reference frame that is consistent with
Gaia-CRF3 to within 0.025 mas yr$^{-1}$ in proper motion and 0.044 mas in position, making it the first ICRS-based reference frame precise enough to
probe the distribution of extended mass within the orbits of stars near SgrA*. In addition, HST-Gaia provides an independent test of the radio measurements
of stellar masers that form the basis of current GC reference frames. We find that the HST-Gaia and radio measurements are consistent to within
0.041 mas yr$^{-1}$ in proper motion and 0.54 mas in position at 99.7\% confidence. Gaia-DR4 is expected to reduce the HST-Gaia reference frame uncertainties
by another factor of $\sim$2x, further improving the reference frame for dynamical studies.
\end{abstract}


\section{Introduction}

The ability to measure stellar motions and orbits at the Milky Way's Galactic center (GC) allows us to study
the central supermassive black hole and surrounding stellar populations at
a level of detail not possible for any other galactic nucleus.
Nearly 30 years of near-infrared (NIR) high-resolution imaging and spectroscopy
of stars in the region have produced a host of groundbreaking discoveries,
including the first conclusive evidence that the central source SgrA* is
a supermassive black hole \citep[SMBH; e.g. ][]{Schodel:2002qq, Ghez:2003ul, Ghez:2008tg, Gillessen:2009to},
testing the effects of General Relativity near an SMBH such as relativistic redshift \citep{GRAVITY-Collaboration:2018cq, Do:2019gr}
and precession \citep{GRAVITY-Collaboration:2020ro, The-GRAVITY-Collaboration:2024xt},
and the discovery of young stars
in an environment where star formation was
thought to be impossible \citep[e.g.][]{Ghez:2003ul, Paumard:2006sh, Lu:2009rq, von-Fellenberg:2022jf, Jia:2023qg}.
A foundational component to these studies is the astrometric
reference frame, which defines the inertial coordinate system
used for the astrometric measurements across these extensive datasets.

An astrometric reference frame is constructed using a set of reference sources with
known positions and motions in a well-defined coordinate system on the sky.
Constructing a reference frame near the GC
is challenging due to the dearth of such reference sources,
especially for observations at NIR wavelengths.
Quasars and other extragalactic sources, which are commonly used
to define reference frames,
cannot be observed near the GC in the NIR due to the large amount
of extinction along the line-of-sight.
The emissive source associated with SgrA* could serve as a reference source
but it is often faint in the NIR and difficult to measure due to
source confusion \citep[e.g.][]{Hornstein:2002gb, Dodds-Eden:2011uk, Weldon:2023la, Paugnat:2024qv}.
One solution that has been adopted is to use stellar SiO masers as reference sources near the GC.
These masers, which are evolved red giant or supergiant stars with extended atmospheres,
are bright sources at both NIR and radio wavelengths.
Since SgrA* is also bright in radio, the positions and motions of the masers can be measured
relative to SgrA* via radio observations.
This establishes a reference frame with a coordinate system
where SgrA* is at rest and located at the origin \citep[e.g.][]{Reid:2007jk, Darling:2023ao}.
Since the masers can also be observed in the NIR,
they can be used as reference sources to establish the astrometric reference frame
for NIR observations \citep{Menten:1997vh, Reid:2003ai, Ghez:2008tg, Gillessen:2009to, Yelda:2010fu, Plewa:2015ud, Sakai:2019fm}.

However, using SiO masers to define the GC reference frame has its own set of challenges.
First, there are a limited number of masers in close proximity of SgrA*, with only 8
known masers within r $\lesssim$ 15".
While additional masers can be found at larger radii,
the field-of-view (FOV) provided by ground-based NIR observations are
restricted due to the technical limitations of the adaptive-optics (AO) corrections
required to achieve high-precision astrometry in the region.
In addition, the radio emission from an SiO maser
originates from an extended circumstellar shell
with a typical radius of $\sim$3 -- 6 AU from the star \citep[e.g.][]{Kemball:2007dr, Gonidakis:2010yt}.
At the GC, such a shell corresponds to an angular size of $\sim$0.4 mas – 0.8 mas,
which is larger than the single-epoch position uncertainty of $\sim$0.2 mas that can be
achieved by radio observations \citep{Darling:2023ao}.
Thus, asymmetries or variability in the shell can cause increased
errors (statistical and/or systematic) in the radio astrometry.

The advent of the Gaia satellite has introduced new opportunities to
improve the reference frame at the GC.
The Gaia-DR3 catalog reports the positions and proper motions
of $\sim$1.5 billion sources across the sky measured in
the Gaia Celestial Reference Frame 3 \citep[Gaia-CRF3;][]{Gaia-Collaboration:2022cm, Gaia-Collaboration:2022sf},
which is defined by extragalactic sources and ultimately tied to the International Celestial Reference System \citep[ICRS;][]{Arias:1995rb}.
As an optical telescope, Gaia does not have the sensitivity required to
measure stars at the distance of the GC. However, it can observe foreground stars along the line-of-sight
that can serve as a new set of reference sources for the GC NIR reference frame.
A challenge of this approach is that the Gaia stars are located at relatively large projected distances from SgrA*
and generally do not overlap with the ground-based AO observations of stars near the SMBH (R $\lesssim$ 25"),
and so they cannot be directly used to establish the reference frame for these observations.
A recent catalog from the VISTA Variables in the Via Lactea (VVV) survey
provides NIR astrometry for stars over a wide area in the Galactic bulge that is tied to Gaia-CRF3 \citep{Griggio:2024dt},
but it is limited in the region surrounding SgrA* (8 stars within R $\lesssim$ 25") and has proper motion
uncertainties of $\gtrsim$0.7 mas yr$^{-1}$, which is insufficient for many high-precision science cases
at the GC.

We use multiple epochs of Hubble Space Telescope (HST) Wide-Field Camera 3 (WFC3-IR) observations of
the GC to bridge the gap between the Gaia reference frame
and the AO observations of stars orbiting SgrA*.
HST has the field-of-view necessary to observe enough Gaia stars to accurately
define the reference frame as well as the sensitivity and spatial resolution to observe stars at the GC.
In this paper, we present a subset
of 2876 stars near SgrA* (R $\lesssim$ 25'', or R $\lesssim$ 1 pc) with positions and proper motions derived from
HST astrometry that has been transformed into Gaia-CRF3.
This catalog provides a sample of astrometric reference stars which defines
the ``HST-Gaia reference frame'' which, unlike past maser-based GC reference
frames, is in the ICRS coordinate system.
A key advantage of HST-Gaia over maser-based reference frames
is that there are significantly more Gaia stars than masers available to define the coordinate system.
In addition, it is the first GC NIR reference frame that is independent
of the radio maser measurements, enabling tests for systematics in
maser-based reference frames.

This paper is organized as follows.
The HST observations and measurements are described
in $\mathsection$\ref{sec:obs}.
The methodology for transforming the HST astrometry into the Gaia-CRF3 reference frame
and deriving stellar proper motions is described $\mathsection$\ref{sec:pm_cat}.
The proper motion catalog of the sample highlighted in this paper is presented
in $\mathsection$\ref{sec:results}, along with an analysis of the consistency
between the HST-Gaia and radio measurements for stellar maser near SgrA*.
The uncertainties in the HST-Gaia reference frame are compared to
previous GC NIR reference frames,
applications of the HST-Gaia reference frame to GC science,
and the expected improvement in the HST-Gaia reference frame with Gaia-DR4
are discussed in $\mathsection$\ref{sec:discussion}.
Finally, our conclusions are summarized
in $\mathsection$\ref{sec:conclusions}.

\section{HST Observations and Measurements}
\label{sec:obs}
\begin{figure}
\begin{center}
\includegraphics[scale=0.4]{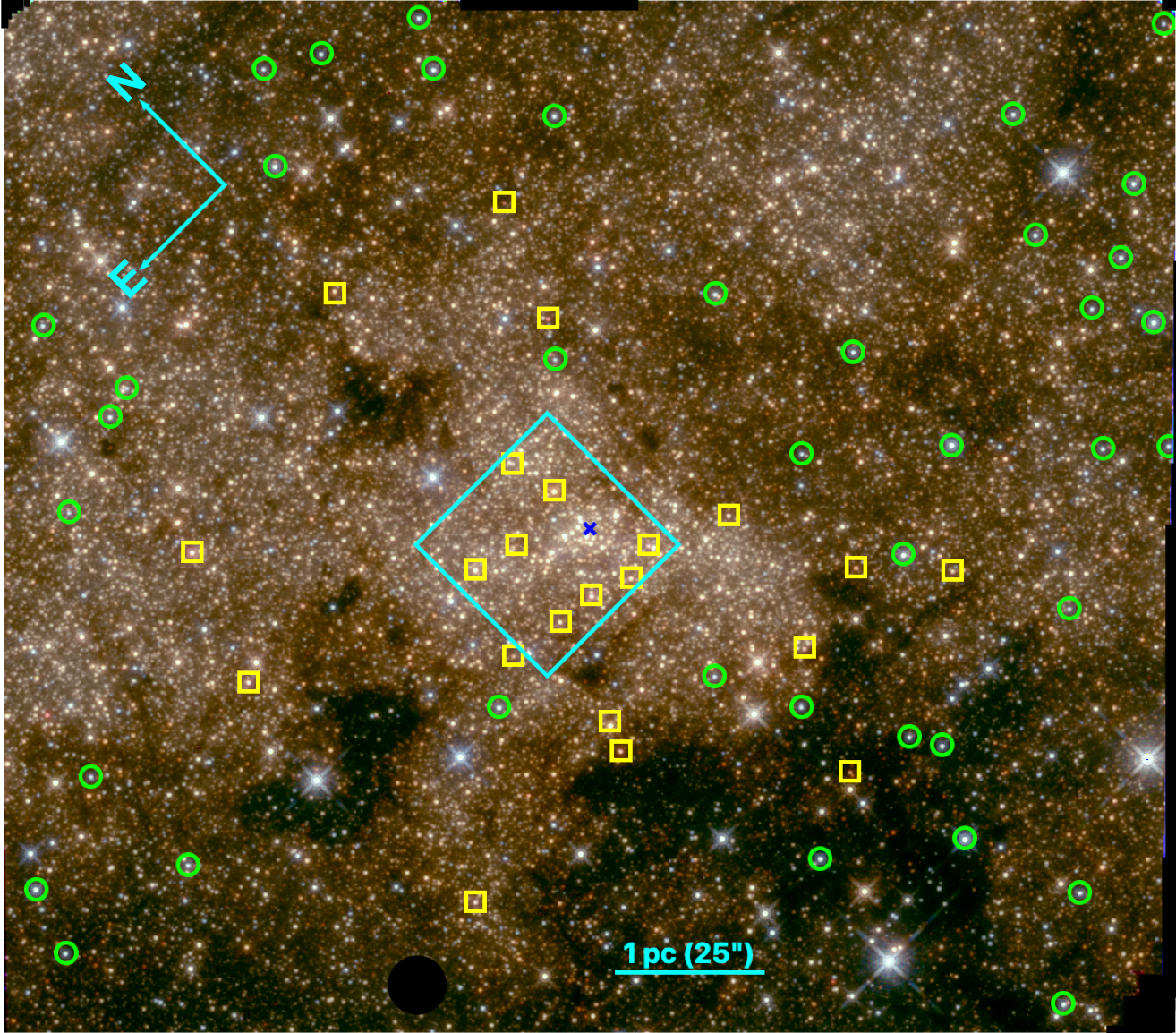}
\caption{The HST WFC3-IR field (red = F153M, green = F139M, blue = F127M), which is approximately centered on SgrA* (blue cross).
The HST astrometry is transformed into Gaia-CRF3 using 40 primary reference stars found in the Gaia-DR3 catalog (green circles).
We report HST-Gaia positions and proper motions for these sources as well as for 2823 secondary reference stars located in a 22"x22" region that
overlaps ground-based AO observations of the GC \citep[cyan box;][]{Sakai:2019fm} and for 13 stellar masers found in the field (yellow squares).
}
\label{fig:img}
\end{center}
\end{figure}

Multiple epochs of observations of a 2$^\prime$ x 2$^\prime$ region centered on SgrA* ($\alpha$(J2000) = 17$^h$45$^m$40.04$^s$, $\delta$(J2000) = -29$^{\circ}$00$'$28$''$.10) were obtained with the the HST WFC3-IR channel between 2010.6261 -- 2023.6178 (Figure \ref{fig:img}).
This dataset contains 14 epochs of observations in the F153M filter for the purpose
of measuring proper motions and a single epoch of observations in the F127M and F139M filters
to measure stellar colors.
The F153M observations were taken over a span of different HST General Observer (GO)
programs and thus have different depths and position angles (PAs).
A summary of the observations is provided in Table \ref{tab:obs},
all of which can be found in the Barbara A. Mikulski Archive for Space Telescopes (MAST)\footnote{\dataset[10.17909/vre6-c497]{http://dx.doi.org/10.17909/vre6-c497}}.

Stellar astrometry and photometry are extracted for each epoch using the
same procedure described in \citet[][]{Hosek:2022af} and references therein.
Initial detection and measurement of stars in each image
are obtained by the \texttt{FORTRAN} routine
\texttt{img2xym\_wfc3ir} \citep[a precursor to the package \texttt{hst1pass} described in][]{Anderson:2022vs}.
The stars are fit using a library of spatially-variable PSFs for the appropriate filter, which are arranged in a 3 x 3 grid that spans the field.
The star images in each exposure are compared to the library PSFs and a uniform perturbation to the PSFs is applied
to minimize the PSF residuals.
A second round of measurements is then performed using the perturbed PSFs to produce a starlist containing star positions and magnitudes.
The starlists for the individual images within each epoch are transformed into a common
reference frame using a first-order polynomial transformation (three free parameters per dimension)
based on the pixel positions of common stars.

A final combined starlist is created for each epoch from the
average astrometry and photometry across the individual images within that epoch.
This is done using
the \texttt{FORTRAN} routine \texttt{KS2} \citep[][]{Anderson:2008qy, Bellini:2017xy, Bellini:2018ow}.
\texttt{KS2} uses the transformations derived from the initial starlists
to combine the images within the epoch into a single average image.
Stars are detected in the average image, where fainter stars can
be detected compared to the individual images alone.
Astrometric and photometric measurements of each star are then made in the individual images
with the neighboring stars subtracted, and then the average positions and fluxes
are computed for each epoch.

The astrometric error of each star is taken to be the error-on-the-mean of
its positions measured in all images within an epoch
($\sigma_{HST}$ = $\sigma_{img}$ / $\sqrt{N_{img}}$,
where $\sigma_{HST}$ is the astrometric error and $\sigma_{img}$ is the standard deviation of the positions across $N_{img}$).
The intrinsic photometric error is taken to be the standard deviation of the
instrumental magnitudes measured across the images, which was found to more accurately reflect
the true photometric scatter than the error-on-the-mean \citep[][]{Hosek:2015cs}.
The instrumental magnitudes are converted into Vega magnitudes using the \texttt{KS2}
zeropoints from \citet{Hosek:2018lr}, and the error on the zeropoints ($\sim$0.01 mags)
is added in quadrature with the intrinsic photometric error.
The final starlists typically contain $\sim$50,000 stars measured in each epoch of the
F153M observations and $\sim$40,000 stars measured in the F127M and F139M observations
across the full HST field.
Note that in this paper we will mainly  focus on sources within R$\leq$25`` (1 pc) of SgrA*;
the measurements across the entire field will be presented in future work (Hosek et al., in prep).

\begin{deluxetable*}{c c c c c c c c}
\tabletypesize{\footnotesize}
\label{tab:obs}
\tablecaption{HST WFC3-IR Observations}
\tablehead{
\colhead{Date} & \colhead{Program ID/PI} &  \colhead{Filter} & \colhead{PA} & \colhead{N$_{img}$} & \colhead{t$_{img}$} & \colhead{t$_{tot}$} & Used for Astrometry? \\
(Year) &  &  & (deg) &  & (s) & (s) &
}
\startdata
2010.6261 & 11671/Ghez & F127M & -45.3 & 12 & 599.23 & 7190 & False \\
2010.6258 & 11671/Ghez & F139M & -45.3 & 10 & 349.23 & 3492 & False \\
2010.4918 & 11671/Ghez & F153M & -45.3 & 21 & 349.23 & 7333 & True \\
2011.6863 & 12318/Ghez & F153M & -45.3 & 21 & 349.23 & 7333 & True\\
2012.6052 & 12667/Ghez & F153M & -45.3 & 21 & 349.23 & 7333 & True\\
2014.099 & 13049/Do & F153M & 134.7 & 4 & 249.23 & 997 & True\\
2018.1064 & 15199/Do & F153M & 134.7 & 16 & 349.23 & 5538 & True\\
2019.162 & 15498/Do & F153M & 134.7 & 16 & 349.23 & 5538 & True\\
2019.6701 & 16004/Do & F153M & -45.3 & 27 & 349.23 & 9429.3 & True\\
2019.7046 & 16004/Do & F153M & -45.3 & 27 & 349.23 & 9429.3 & True\\
2019.7924 & 16004/Do & F153M & -45.3 & 27 & 349.23& 9429.3 & True\\
2020.2234 & 15894/Do & F153M & 134.7 & 20 & 349.23 & 6985 & True\\
2022.1428 & 16681/Do & F153M & 134.7 & 21 & 299.23 & 6284 & True\\
2022.5441 & 16681/Do & F153M & -45.3 & 21 & 299.23 & 6284 & True\\
2023.1535 & 16990/Do & F153M & 134.7 & 21 & 299.23 & 6284 & True\\
2023.6178 & 16990/Do &  F153M & -45.3 & 21 & 299.23 & 6284 & True\\
\enddata
\tablecomments{Description of columns. Date: average date of observations (calculated assuming the Gregorian calendar), Program ID/PI: HST program ID and PI, Filter: filter, PA: position angle, N$_{img}$: number of images, t$_{img}$: integration time per image; t$_{tot}$: total integration time.}
\end{deluxetable*}

\section{HST-Gaia Proper Motions}
\label{sec:pm_cat}
We measure stellar proper motions by transforming each epoch of HST observations
into the Gaia-CRF3 reference frame \citep{Gaia-Collaboration:2022cm}.
These transformations are calculated using stars in common between
HST and Gaia-DR3. We refer to these as the primary reference stars,
which are identified in $\mathsection$\ref{sec:gaia_ref}.
Stellar proper motions are then measured
using a new method that utilizes Gaussian Processes to model systematic errors as
described in $\mathsection$\ref{sec:pm_catalog}.
We refer to these measurements as the HST-Gaia proper motions.
The consistency between the HST-Gaia reference frame established
by these measurements and Gaia-CRF3 is then assessed in $\mathsection$\ref{sec:ref_frame_comp}.

The Gaia-CRF3 reference frame is a realization of the ICRS, which defines
a coordinate system that is defined relative to extragalactic objects with no proper motion \citep{Arias:1995rb}.
By extension, the HST-Gaia proper motions are in the ICRS coordinate system as well.
However, for some applications it is useful to measure stellar proper motions in a
coordinate system where SgrA* is at the origin and has no proper motion\footnote{Note that in SgrA* has a non-zero proper motion in the ICRS due to the reflex motion of the Sun's orbit around the Galaxy. This motion is defined observationally.},
which we refer to as SgrA*-at-Rest coordinates.
We discuss how to convert the HST-Gaia proper motions
into SgrA*-at-Rest coordinates in $\mathsection$\ref{sec:sgra_rest}.
We refer to these as the SgrA*-at-Rest proper motions.

\subsection{Transforming HST Astrometry into the Gaia-CRF3 Reference Frame}
\label{sec:gaia_ref}
The HST astrometry is transformed into the Gaia-CRF3 reference frame
using a set of primary reference stars that are well-measured in both the HST and Gaia-DR3 catalogs.
However, there are a limited number of such sources available.
Due to the high extinction near the GC \citep[A$_{Ks}$ $\sim$ 2.7 mag;][]{Schodel:2010eq},
most of the NIR sources detected by HST are too faint at optical
wavelengths to be detected by Gaia.
On the other hand, sources detected by Gaia must come from the less-extinguished
foreground population and may be saturated in the HST data.
In addition, the high stellar crowding in the field can bias
the astrometric measurements of either Gaia or HST.
Thus, the selection of the primary reference stars
must be made with care.

To begin, we make a series of quality cuts to the Gaia-DR3 catalog to identify
Gaia sources in the region with reliable astrometry \citep[][]{Arenou:2018dz, Lindegren:2021ae, Fabricius:2021yo}.
We consider Gaia-DR3 stars with the following catalog values:

\begin{itemize}
\item \texttt{astrometric\_params\_solved} = 31, indicating that the source has a five-parameter astrometric solution (position, parallax, and proper motion) and that accurate color-dependent PSF correction terms can be applied

\item  \texttt{duplicated\_source} = False, which requires that no other \emph{Gaia} source is detected within 0.$''$18

\item \texttt{parallax\_over\_error} $\geq$ -10, eliminating sources with significantly unphysical parallaxes

\item \texttt{astrometric\_excess\_noise\_sig} $\leq$ 2, removing sources with significant excess noise in the Gaia astrometric solution beyond the statistical errors

\item \texttt{RUWE} $\leq$ 1.4, also to remove sources with significant excess noise relative to the Gaia astrometric solution

\item \texttt{phot\_g\_mean\_mag} $>$ 13 mag, to avoid additional systematic errors in the Gaia astrometry for bright sources

\end{itemize}

There are 55 Gaia-DR3 stars that overlap with the HST field after these cuts.
The positions of the Gaia stars are converted into angular offsets from
a fiducial position near the center of the HST field, chosen to be
($\alpha_f$(ICRS), $\delta_f$(ICRS)) = (17$^h$45$^m$40.032863$^s$, -29$^{\circ}$00$'$28$''$.24260).
These offsets are denoted as $\delta_{\alpha*}$ = ($\alpha$ - $\alpha_{f}$)cos($\delta$)
and $\delta_{\delta}$ = ($\delta$ - $\delta_f$).

The final set of primary reference stars is selected using an iterative process.
First, the positions
of the Gaia stars
at each HST epoch are calculated using the Gaia-DR3 proper motions,
with the uncertainties in the Gaia positions calculated
via the propagation of Gaia catalog errors.
Then, the Gaia stars are matched to HST sources,
and Gaia stars with corresponding HST sources with F153M $\leq$ 13 mag are removed
as a conservative limit to avoid saturation in the HST image.
This eliminates 5 Gaia stars from the sample.
Next, the remaining Gaia stars are used to calculate a
second-order polynomial via linear least squares minimization
to transform HST pixel coordinates (x, y)
into Gaia-CRF3 coordinates ($\delta_{\alpha*}$, $\delta_{\delta}$):
\begin{equation}
\label{eq:transx}
\delta_{\alpha*} = a_0 + a_1x + a_2y + a_3xy + a_4x^2 + a_5y^2
\end{equation}
\begin{equation}
\label{eq:transy}
\delta_{\delta} = b_0 + b_1x + b_2y + b_3xy + b_4x^2 + b_5y^2
\end{equation}
where $\{a_0, a_1, ..., b_0, b_1, ...\}$ are the coefficients of the polynomials.
A second-order polynomial is used because it was found to be the most effective
as reducing the residuals to the fit. Testing revealed that a first-order polynomial
resulted in statistically significant position residuals for many of the reference stars,
while a third-order polynomial did not not offer a noticeable improvement over the second-order polynomial.

Next, the uncertainty in the transformation ($\sigma_{trans}$)
is characterized using a full sample bootstrap over the reference stars.
For each epoch, the reference stars are resampled with replacement for
100 iterations.
The transformations are recalculated for each iteration and applied to HST astrometry.
For each star, the standard deviation of the transformed positions across the iterations is
taken to be $\sigma_{trans}$ for that star.
The total astrometric error ($\sigma_{ast}$) for a star in a given epoch
is then calculated as:
\begin{equation}
\label{eq:ast_err}
\sigma_{ast} = \sqrt{\sigma_{HST}^2 + \sigma_{trans}^2}
\end{equation}
where $\sigma_{HST}$ is the astrometric error of the HST measurements (see $\mathsection$\ref{sec:obs}).
The HST-Gaia proper motions of the stars are then calculated from the transformed
HST astrometry and corresponding $\sigma_{ast}$ values using the methodology described in $\mathsection$\ref{sec:pm_catalog}.

Lastly, the sample of Gaia reference stars is examined for outliers by
comparing their HST-Gaia proper motions
to their original values from the Gaia-DR3 catalog.
Reference stars with either position or proper motion differences larger than 5$\sigma$ (where $\sigma$ is the quadratic sum
of the HST-Gaia and Gaia-DR3 errors) are considered outliers.
This conservative threshold is adopted to identify the strongest outliers in the sample, which
would have undue influence on the transformation parameters.
If one or more outliers are present, then the largest outlier is eliminated from the
reference star sample and the process is repeated.
These iterations continue until no outliers are found to exist, which was achieved
after 10 Gaia stars are eliminated from the primary reference star sample.
Many of the Gaia stars removed in this manner were found
to either lie within several pixels of the edge of the HST field (visibly truncating their PSF)
or be in close proximity to neighboring stars, compromising their HST astrometry.

\begin{figure*}
\begin{center}
\includegraphics[scale=0.35]{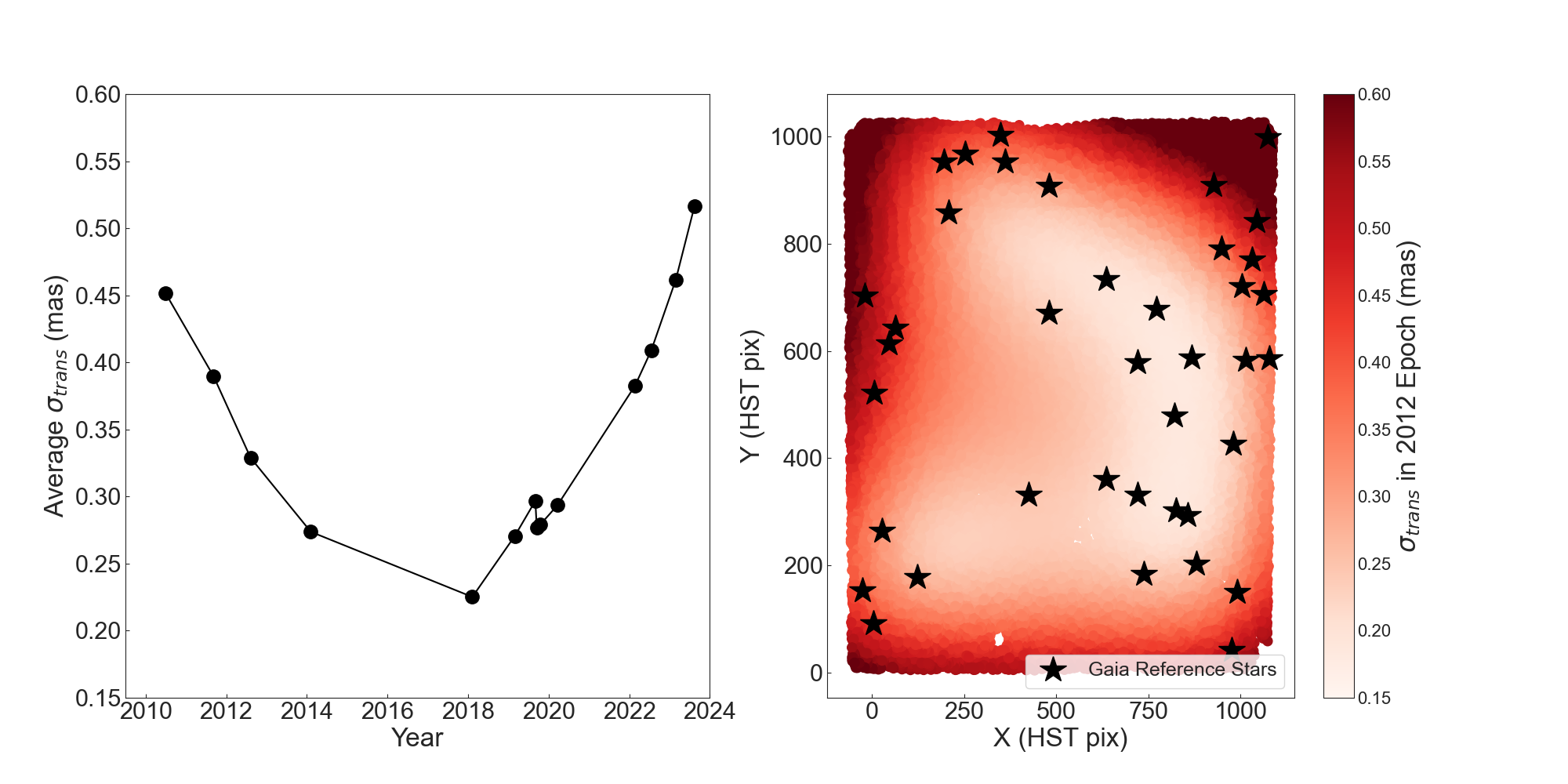}
\caption{The transformation errors ($\sigma_{trans}$) for the HST-Gaia astrometry, which varies as a function of time and position.
Left: Average $\sigma_{trans}$ as a function of time, which generally increases as the time from the Gaia-DR3 reference epoch (2016) increases.
Right: $\sigma_{trans}$ as a function of position on the HST detector for the 2012 epoch (pixel scale = 1.21" pix$^{-1}$) , with the positions of the Gaia reference stars
shown by black stars. $\sigma_{trans}$ generally increases near the edges of the field as well as in areas with fewer reference stars.}
\label{fig:trans_err}
\end{center}
\end{figure*}

The final set of primary reference stars used to transform the HST astrometry
into the Gaia-CRF3 reference frame is comprised of 40 sources.
These stars span a range of Gaia magnitudes between 15.1 mag $<$ G $<$ 20.2 mag and have
a median position error of 0.15 mas and a median proper motion error of 0.16 mas yr$^{-1}$ in the Gaia DR3 catalog (Table \ref{tab:gaia_orig}).
The average $\sigma_{trans}$ achieved by the resulting transformations
varies between 0.2 -- 0.5 mas depending on the epoch (Figure \ref{fig:trans_err}).
The fact that $\sigma_{trans}$ generally increases as
the time from the Gaia-DR3 reference epoch (2016.0) increases
indicates that
the transformation errors are
limited by the Gaia-DR3 measurement errors for the reference stars.
Figure \ref{fig:trans_err} also shows how $\sigma_{trans}$ varies
as a function of position in the HST field,
with increasing error towards the field edges and in regions where there are
fewer Gaia reference stars nearby.

\movetableright=3mm
\begin{deluxetable*}{lccccccccc}
\tablewidth{0pt}
\tabletypesize{\tiny}
\tablecaption{Primary Reference Stars: Gaia DR3 Catalog Measurements}
\tablehead{
\colhead{source\_id} & \colhead{phot\_g\_mean\_mag} & \colhead{ra} & \colhead{ra\_error} & \colhead{dec} & \colhead{dec\_error} &
\colhead{pmra} & \colhead{pmra\_error} & \colhead{pmdec} & \colhead{pmdec\_error} \\
& (mag) & (deg) & (mas) & (deg) & (mas) & (mas/yr) & (mas/yr) & (mas/yr) & (mas/yr)
}
\startdata
4057481683573622528 & 18.24 & 266.39940889 & 0.173 & -29.01953897 & 0.138 & 2.050 & 0.202 & -4.857 & 0.130 \\
4057481683573626880 & 15.88 & 266.39651474 & 0.043 & -29.01633561 & 0.033 & 0.263 & 0.049 & -2.250 & 0.031 \\
4057481717933348608 & 17.39 & 266.41697834 & 0.101 & -29.02998491 & 0.080 & -6.113 & 0.118 & -5.658 & 0.073 \\
4057481717933350528 & 17.99 & 266.41359224 & 0.149 & -29.02779105 & 0.120 & -7.100 & 0.173 & -5.444 & 0.108 \\
4057481717933358848 & 17.25 & 266.41955472 & 0.094 & -29.02096452 & 0.072 & 1.789 & 0.110 & 0.788 & 0.066 \\
4057481717933359360 & 18.04 & 266.41333717 & 0.147 & -29.02119722 & 0.119 & 0.019 & 0.174 & -5.168 & 0.111 \\
4057481717934037760 & 19.12 & 266.41807115 & 0.326 & -29.03107056 & 0.258 & 3.333 & 0.371 & 1.031 & 0.229 \\
4057481717938509312 & 18.59 & 266.41397518 & 0.198 & -29.02023107 & 0.159 & -1.315 & 0.246 & -3.164 & 0.152 \\
4057481722235384320 & 15.12 & 266.41521387 & 0.029 & -29.02387599 & 0.024 & -0.223 & 0.036 & -1.737 & 0.022 \\
4057481722235386368 & 18.56 & 266.41602804 & 0.201 & -29.01701274 & 0.158 & 1.363 & 0.239 & 2.421 & 0.147 \\
\enddata
\tablecomments{Columns are as defined in the Gaia DR3 source catalog. All positions are at the Gaia DR3 reference epoch of 2016.0.}
\tablecomments{This table is available in its entirety in the machine-readable format.}
\label{tab:gaia_orig}
\end{deluxetable*}

\subsection{Modeling Proper Motion Systematics using Gaussian Processes}
\label{sec:pm_catalog}
Systematic errors can introduce correlations between astrometric measurements that are
not related to the proper motion of a given source.
Possible sources of systematic errors include stellar crowding,
where a neighboring star moves close enough to a given star
to bias its astrometry \citep[e.g.][]{Jia:2019oq},
and shifts in the astrometric reference frame, which would
introduce correlated astrometric residuals as a function of time.
We find an additional systematic in the HST astrometry
that causes the star positions to shift depending on
the position angle (PA) of the observations.
The size of the position shift increases with magnitude,
reaching as large as $\sim$2 mas for faint stars (see Appendix \ref{app:ast_errs}).
These systematic error must be addressed in order
to accurately measure proper motions.

We introduce a new approach to model proper motions in the presence of such systematics.
We utilize Gaussian Processes \citep{Rasmussen:2006zk}
to simultaneously fit stellar proper motion and systematic correlations.
These models use one or more kernels (see Appendix \ref{app:gpoly}):
\begin{itemize}
\item \emph{Polynomial kernel} ($K^{\mathrm{(poly)}}$): Fits the motion of the star as a function of time with a polynomial. We use a first-order polynomial to measure the linear proper motion of a star:
\begin{equation}
\label{eq:pm}
x' = x_0 + \mu_x(t - t_0)
\end{equation}
where $x'$ is the observed position at time $t$ (i.e. either $\delta_{\alpha^*}$ or $\delta_{\delta}$),
$x_0$ is the $x$ position at $t_0$, $\mu_x$ is the proper motion,
and $t_0$ is the astrometric-error weighted average time across the data (Appendicies \ref{app:kernels:poly}, \ref{app:poly_higher_order_pred}).

\item \emph{Squared-exponential kernel} ($K^{\mathrm{(sqexp)}}$): Models time-correlated data using a squared-exponential function.
We use this to model systematics that are induced by shifts in the astrometric reference frame or observing PA (Appendix \ref{app:kernels:timecorr}).

\item \emph{Step kernel} ($K^{\mathrm{(step)}}$): Models time-correlated data using a step function.  We use this as an alternative kernel to capture systematics induced by shifts in the astrometric reference frame or observing PA (Appendix \ref{app:kernels:timecorr}).

\item \emph{Confusion kernel} ($K^{\mathrm{(conf)}}$): Models systematics correlated with spatial position, such as biases
induced by stellar confusion with a nearby source (Appendix \ref{app:kernels:spatial}).

\end{itemize}
In addition, an additive error can be included in the model which inflates the measurement uncertainties for each data point by a constant value. This is useful in the case where the measurement uncertainties are systematically underestimated.

Before applying these kinematic models to derive stellar proper motions,
we remove severe astrometric outliers from the data
via leave-one-out (LOO) cross-validation \citep[e.g.][see also Appendix \ref{app:loocv}]{Vehtari:2015qa, Do:2019gr}.
Such outliers can be caused by cases of incorrect source matching between epochs or significant stellar confusion.
For each star, we fit an initial kinematic model comprised of only the polynomial kernel
to the data (the $\alpha^*$ and $\delta$ positions are fit independently).
If the difference between an observed data point and its
predicted value from a fit with that point removed
(i.e., the LOO prediction)
is greater than 5x the uncertainty in the
LOO prediction in either $\alpha^*$ or $\delta$, then that data point is
identified as an outlier and removed (e.g. a 5$\sigma$ outlier cut, as a conservative threshold
for identifying outliers).
This process removes $\sim$5.5\% of the total number of
HST astrometric measurements from the sample.

After outlier rejection, proper motions are derived for all stars that have at least $3$
epochs of F153M astrometry.
In cases where are $\leq$4 epochs of astrometry,
then there aren't enough data to constrain the kinematic models with
additional kernels beyond the polynomial kernel (since each kernel has $2$ free parameters).
Instead, linear regression is used to fit the proper motions according to Equation \ref{eq:pm}.
For sources with $\geq$ 5 epochs of astrometry,
several kinematic models are fit to the data:
\begin{itemize}
\item \emph{poly-only}: polynomial kernel ($K^{\mathrm{(poly)}}$) only
\item \emph{poly+sqexp}: polynomial kernel ($K^{\mathrm{(poly)}}$) plus the squared-exponential kernel ($K^{\mathrm{(sqexp)}}$).
\item \emph{poly+step}: polynomial kernel ($K^{\mathrm{(poly)}}$) plus the step kernel ($K^{\mathrm{(step)}}$).
\item \emph{poly+confusion}: polynomial kernel ($K^{\mathrm{(poly)}}$) plus the confusion kernel ($K^{\mathrm{(conf)}}$)
\item \emph{poly+add}: polynomial kernel ($K^{\mathrm{(poly)}}$) plus an additive error
\end{itemize}
For the sample reported in this paper, only one star has $\leq$4 epochs of astrometry and is fit using linear
regression while the rest have $\geq$ 5 epochs and are fit using these kinematics models.

We use the Expected Log Probability Distribution \citep[ELPD; see][]{Vehtari_2016, Gelman_2013} to determine which
kinematic model is preferred:
\begin{equation}
\label{eq:elpd}
\mathrm{ELPD} =  \sum_i^N \log(P(x_i \mid \mat{x}_{\minus i}))
\end{equation}
where $P(x_i \mid \mat{x}_{\minus i})$ is the probability of observing the $i$-th data point given
the model that excludes this data point and $N$ is the number of data points in the fit.
If the model is a Gaussian Process then $x_i$ is distributed normally:
\begin{equation}
x_i \sim \mathcal{N}\left(x_{\mathrm{loo}_i},  \sigma^2_{\mathrm{loo}_i}\right)
\end{equation}
where $x_i$ is the observed value of the $i$-th data point,
$x_{\mathrm{loo}_i}$ is the LOO prediction for the value of that data point,
and $\sigma^2_{\mathrm{loo}_i}$ is the variance of the LOO prediction, which  (see Appendix \ref{app:loocv}).
Thus:
\begin{equation}
\label{eq:loo_prob}
P(x_i \mid \mat{x}_{\minus i}) = \frac{1}{\sqrt{2\pi\sigma^2_{\mathrm{loo}_i}}} \exp{\left[-\frac{(x_i - x_{\mathrm{loo}_i})^2}{2\sigma^2_{\mathrm{loo}_i}}\right]}
\end{equation}
The ELPD is calculated for the $\alpha^*$ and $\delta$ fits and then are summed
together to calculate the total ELPD for a given model of a given star.
Note that the ELPD will favor a model that best describes the variance in the data
(i.e., minimizes $x_i - x_{\mathrm{loo}_i}$ relative to $\sigma^2_{\mathrm{loo}_i}$)
with the lowest possible LOO uncertainties possible.
Further, the ELPD penalizes models that overfit the data through the use of the LOO prediction:
for overfit models, $x_i - x_{\mathrm{loo}_i}$ (hereafter referred to as the LOO residual)
will be large, since
the removal of a given data point will significantly change the model.

\begin{figure*}
\begin{center}
\includegraphics[width=0.95\textwidth]{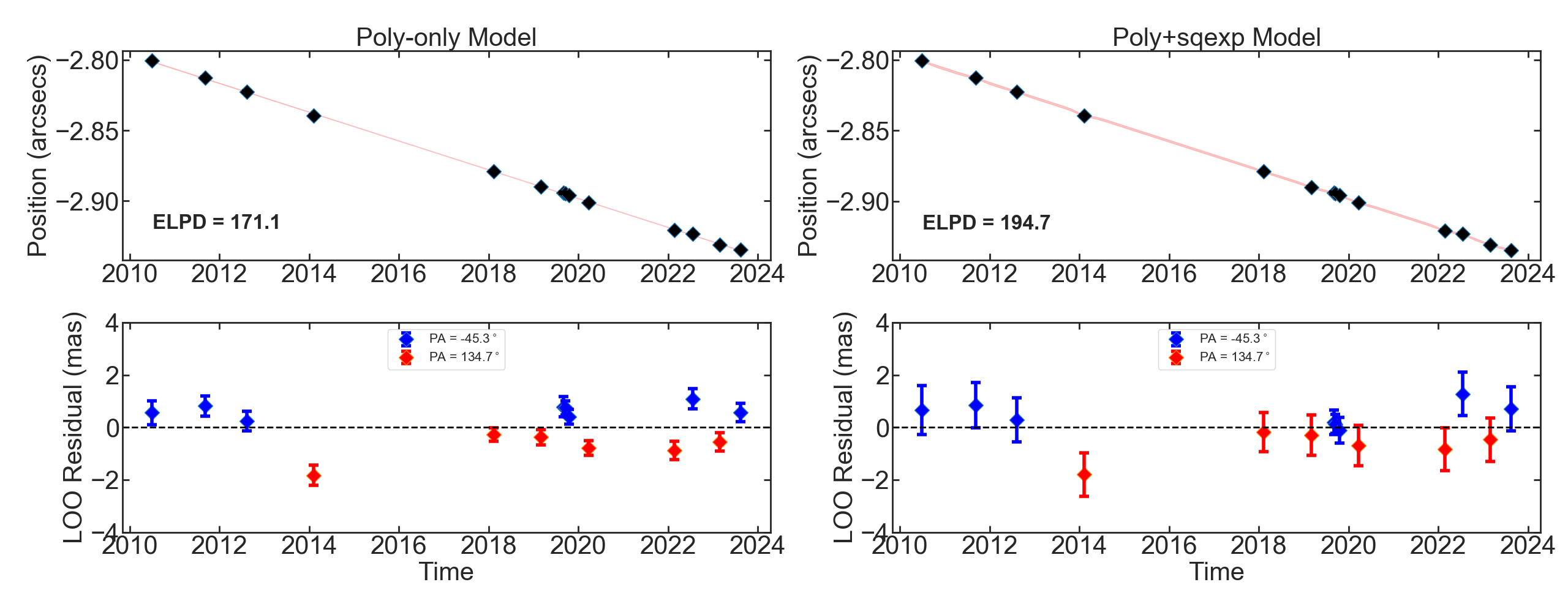}
\caption{An example application of the new kinematic models introduced in this paper to the
proper motion of the star S3-374.
The left panel shows the \emph{poly-only} model fit while the right panel show the \emph{poly+sqexp} model fit.
For each panel, the top plot shows the model (red, with the shaded region showing the uncertainty) compared to the observed data (black points),
with the corresponding ELPD of the model printed to the lower left of the plot.
The bottom plot in each panel shows the LOO residuals between the model and data, where the color of each data point corresponds to the PA
of the epoch and the errorbars represent the LOO uncertainty at each point.
The \emph{poly-only} model shows correlated residuals due to the changing PA of the observations (see Appendix \ref{app:ast_errs}), which produces a larger variance
in the data then can be explained by the model.
These residuals are reduced and better captured by the uncertainty in the \emph{poly+sqexp} model, which is strongly preferred by the ELPD.}
\label{fig:gpoly}
\end{center}
\end{figure*}

We accept a model with two kernels over the single-kernel \emph{poly-only}
model if the ELPD of the two-kernel
model is greater by $\geq6$.
This criterion implies that the probability of the \emph{poly-only} model
being preferred over the two-kernel model is $\sim 0.0025$,
{\em e.g.}, the two-kernel model is preferred at the $\sim3\sigma$ probability level.
If this threshold is met, then we conclude that the additional free parameters
introduced by the two-kernel model are necessary.
The two-kernel model with the
highest ELPD (i.e. the highest probability) is then adopted as the preferred model.
An example where the \emph{poly+sqexp} model is identified as the preferred model
is shown in Figure \ref{fig:gpoly}.

\subsubsection{Assessing the Best-Fit Kinematic Models}
\label{sec:ref_star_properties}

Ideally, the best-fit kinematic model for a given star would provide LOO predictions for
the data that are consistent with the observed values within the LOO uncertainties.
To test this, we define a metric similar to the reduced chi-squared:
\begin{equation}
\label{eq:chi2_red}
\widetilde{\chi^2_{\mathrm{loo}}} = \frac{\chi^2_{\mathrm{loo}}}{\mathrm{N}},
\end{equation}
where $\mathrm{N}$ is the number of astrometric measurements and
\begin{equation}
\label{eq:chi2}
\chi_{\mathrm{loo}}^2 = \sum_{i=1}^{i=N} \frac{(x_{i} - x_{\mathrm{loo}_i})^2}{\sigma^2_{\mathrm{loo}_i}}
\end{equation}
where $x_{i}$ is the observed star position at the $i$th of $N$ astrometric epochs,
$x_{\mathrm{loo}_i}$ is the LOO prediction for the star position from the kinematic model at that epoch,
and $\sigma_{\mathrm{loo}_i}$ is the uncertainty of the LOO prediction (see Appendix \ref{app:loocv}).
This metric provides a measure of the size of the LOO residual for each epoch relative to the LOO uncertainty at that epoch.

For the sample of stars presented in this paper,
we calculate the expected distribution of $\widetilde{\chi^2_{loo}}$ in the ideal case where
the LOO residual are indeed consistent with the LOO uncertainty at each data point.
First, we assume that
the LOO predictions from the best-fit kinematic models represent the true positions of the stars at each of the HST epochs.
Each position is then perturbed by the measurement uncertainty
(i.e., modified by a random value drawn from a Gaussian distribution with a mean of zero and a standard deviation equal
to the observed astrometric error from Equation \ref{eq:ast_err}) to represent a measurement of that star position.
The resulting $\widetilde{\chi^2_{loo}}$ distribution across the predicted measurements is then calculated.
This process is repeated 1000 times to calculate the average
expected distribution for $\widetilde{\chi^2_{\mathrm{loo}}}$
for the sample.

\begin{figure*}
\begin{center}
\includegraphics[scale=0.35]{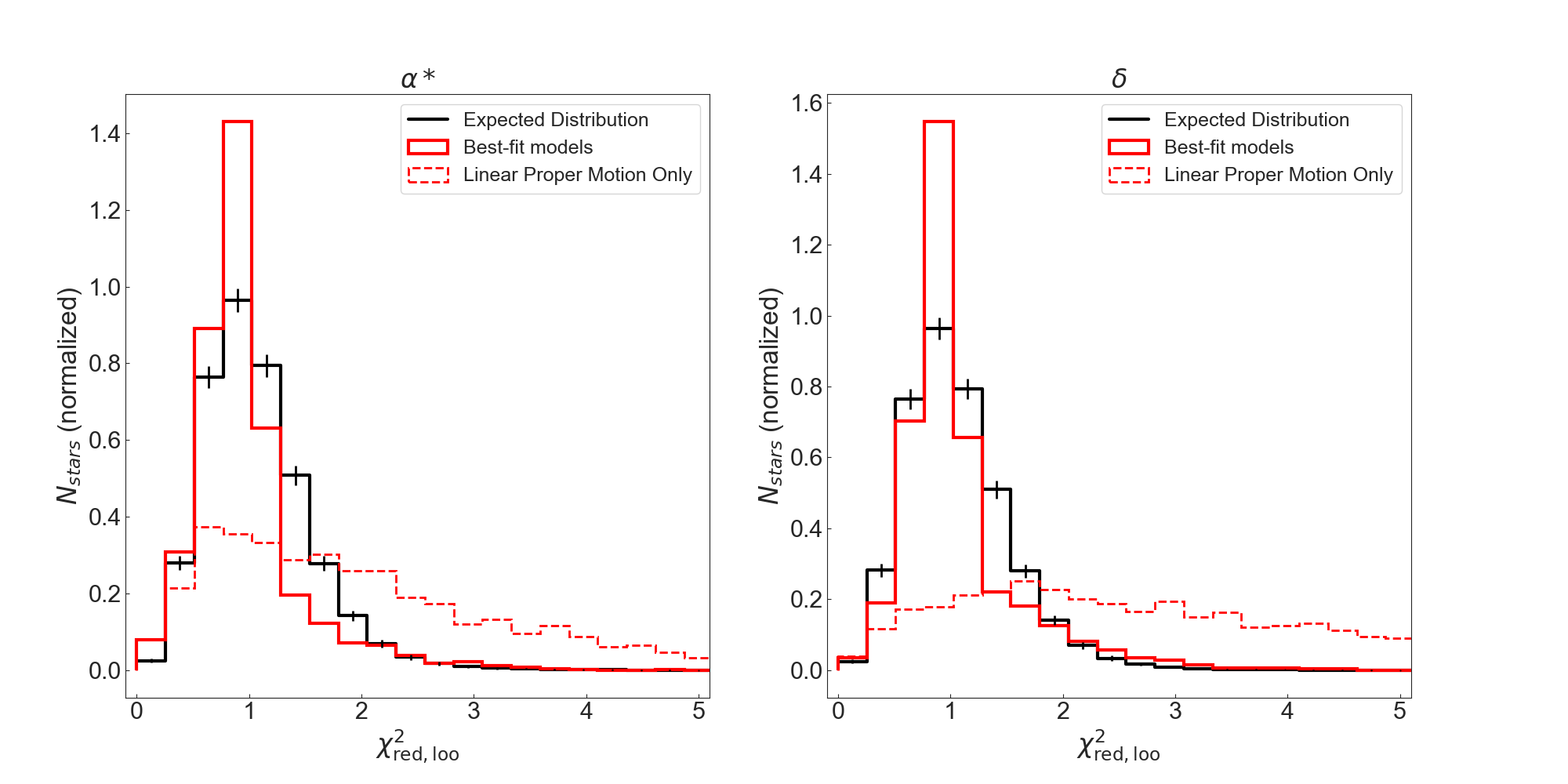}
\caption{The observed and expected distributions of $\widetilde{\chi^2_{\mathrm{loo}}}$ for the astrometric reference stars presented in this paper,
with $\alpha*$ on the left and $\delta$ on the right. In both cases, the observed distribution of the best-fit kinematic models
(solid red histograms) peak at similar values as the expected distributions (black histograms).
However, if \emph{poly-only} models used for all stars,
then the resulting distribution of $\widetilde{\chi^2_{\mathrm{loo}}}$ extends to significantly higher values
then expected due to systematic errors in the astrometry beyond what the \emph{poly-only} models can capture (red dashed histograms).}
\label{fig:chi2_red}
\end{center}
\end{figure*}

If the best-fit kinematic models provide a good description of the data, then
the observed distribution of $\widetilde{\chi^2_{\mathrm{loo}}}$ would
be similar to the expected distribution of $\widetilde{\chi^2_{\mathrm{loo}}}$ for the sample.
This comparison is shown in Figure \ref{fig:chi2_red}.
We find that the peaks of the observed and expected
$\widetilde{\chi^2_{\mathrm{loo}}}$ distributions are indeed quite similar,
with the observed distribution being somewhat narrower than expected.
On the other hand, if \emph{poly-only} models are used for all
of the stars, then the resulting $\widetilde{\chi^2_{\mathrm{loo}}}$ distribution is
significantly broader then the expected distribution.
This indicates that the observed LOO residuals are generally larger than
the uncertainty in the \emph{poly-only} models, suggesting
that there are systematic correlations
in the astrometry that are not captured when only the \emph{poly-only} models are used.
In contrast, the best-fit kinematic models better capture these correlations and generally provide
good fits to the data.

\subsection{The HST-Gaia Reference Frame: Consistency with Gaia-CRF3}
\label{sec:ref_frame_comp}
The HST-Gaia measurements define the ``HST-Gaia reference frame'',
i.e., a reference frame defined by HST measurements that are transformed into Gaia-CRF3.
The transformation errors ($\sigma_{trans}$) calculated in $\mathsection$\ref{sec:gaia_ref} quantify the
statistical error in the transformation into the Gaia-CRF3 reference
frame (e.g., due to the precision of the reference star measurements)
as well as systematic errors that may be present between the reference
stars themselves (e.g., if a particular reference star is biased
relative to the rest of the sample).
Here we assess the overall consistency between the HST-Gaia and
Gaia-CRF3 coordinate systems.
A discrepancy would result in an average systematic bias
between the HST-Gaia and Gaia-DR3 measurements
of the primary reference stars.
The size and uncertainty of this bias represents the highest possible
accuracy and precision of measurements in the HST-Gaia reference frame
relative to the Gaia-CRF3 coordinate system.

We define the bias between the HST-Gaia and Gaia-CRF3
reference frames as the
error-weighted mean difference between the positions and
proper motion values for the primary reference stars:
\begin{equation}
\label{eq:wmean}
\Delta x = \frac{\sum_i^N{w_i (x_{i, HST-Gaia} - x_{i, DR3})}}{\sum_i^N{w_i}}
\end{equation}
where $x_{i, HST-Gaia}$ and $x_{i, DR3}$ are the HST-Gaia and Gaia-DR3 values
for the $i$th reference star, respectively, and the weight
$w_i$ $=$ $1 / \sigma_i^2$, where
$\sigma_i$ is the quadratic sum of the associated HST-Gaia and
Gaia-DR3 errors on those values.
The uncertainty in the bias is then the error on the weighted mean:
\begin{equation}
\label{eq:wEOM}
\sigma_{\Delta x} = \frac{1}{\sqrt{\sum_i^N{w_i}}}
\end{equation}

Calculating the proper motion bias between the HST-Gaia and Gaia-DR3 measurements
is straightforward since the proper motions can simply be subtracted to find the difference.
In order to calculate the position bias we must first use the HST-Gaia and Gaia-DR3
proper motions to infer the positions of the reference stars at a common epoch.
This epoch ($t_{0,comp}$) is chosen to be the error-weighted average reference epoch
of the proper motion measurements:

\begin{equation}
\label{eq:t0_comp}
t_{0, comp} = \frac{\sum^N_i \sigma_{H,i}^2 t_{0,H,i} + \sigma_{G,i}^2 t_{0, G,i} }{\sum^N_i \sigma_{H,i}^2 + \sigma_{G,i}^2}
\end{equation}

where $\sigma_{H,i}$ is the HST-Gaia proper motion error
for the $i$th star, $t_{0,H,i}$ is the HST-Gaia reference epoch for the $i$th star,
and $\sigma_{G,i}$ and  $t_{0, G,i}$ are the analogous values for the Gaia proper motion of the $i$th star.
This represents the time at which the uncertainty in the difference between the HST and Gaia positions
is minimized \citep[e.g.][]{Yelda:2010fu}, which is calculated to be 2016.0605 for this sample.
Then, Equations \ref{eq:wmean} and \ref{eq:wEOM} can be applied using the inferred positions at $t_{0,comp}$.

\begin{figure*}
\begin{center}
\includegraphics[scale=0.3]{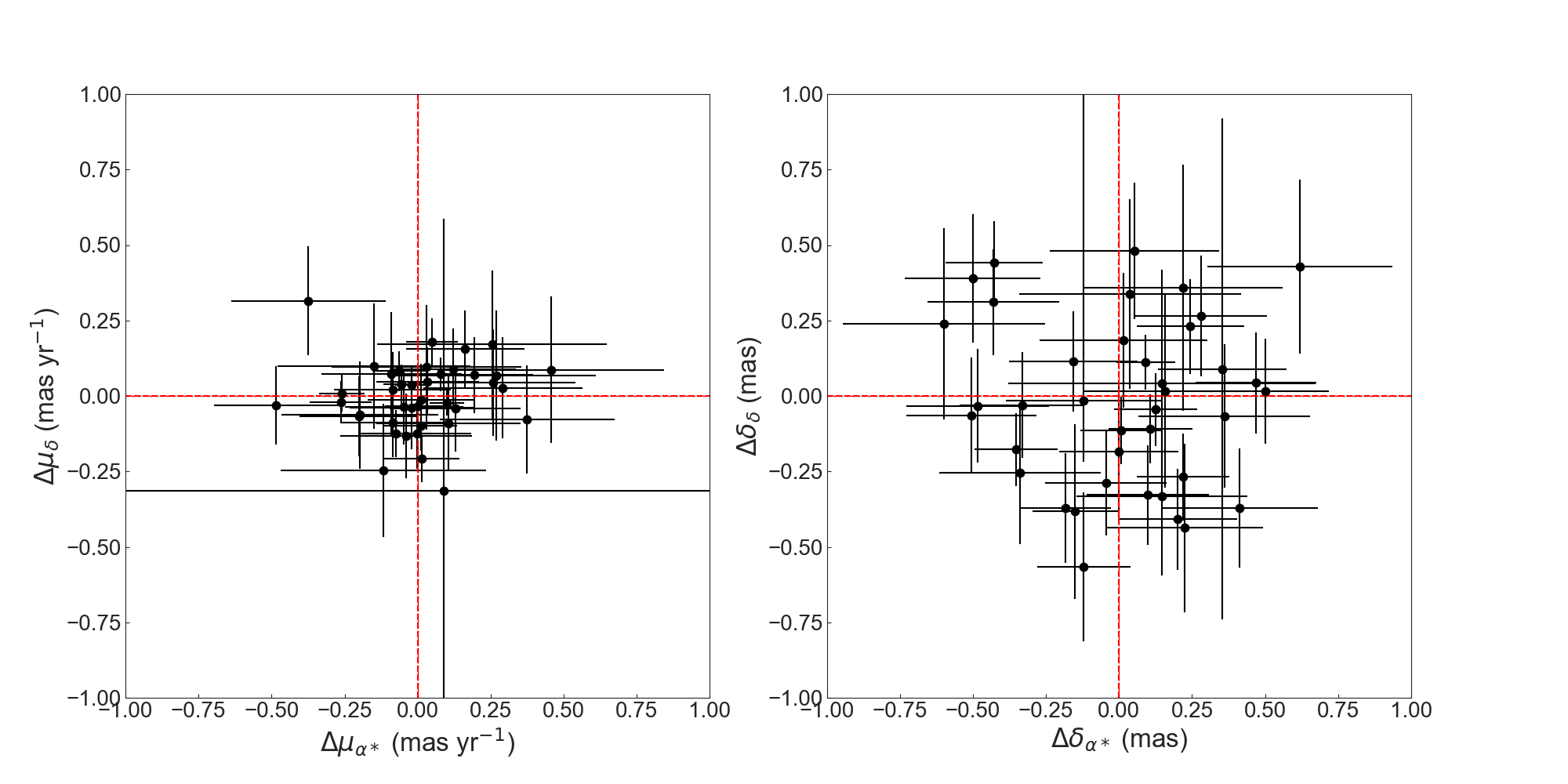}
\caption{The differences between the the HST-Gaia and Gaia-DR3 proper motions (left) and positions (right) for the primary reference stars,
which constrain the consistency between the HST-Gaia and Gaia-CRF3 coordinate systems.
We find no evidence for systematic biases between the reference frames to a precision of 0.025 mas yr$^{-1}$ in proper motion and 0.044 mas in position.
}
\label{fig:pm_pos_comp}
\end{center}
\end{figure*}

The differences between the HST-Gaia and Gaia-CRF3
position and proper motion measurements for the primary reference stars are shown in Figure \ref{fig:pm_pos_comp}.
For the proper motions, we derive the bias to be
$\Delta$$\mu_{\alpha^*}$ = -0.015 $\pm$ 0.020 mas yr$^{-1}$ and
$\Delta$$\mu_{\delta}$ = 0.002 $\pm$ 0.015 mas yr$^{-1}$.
This indicates that there is no evidence for a bias between the HST-Gaia and Gaia-CRF3 proper
motions to a total precision of
$\sigma_{\Delta pm}$ = $\sqrt{\sigma_{\Delta \mu_{\alpha*}}^2 + \sigma_{\Delta \mu_{\delta}}^2}$ = 0.025 mas yr$^{-1}$.
For the positions, we find ($\Delta$$\delta_{\alpha^*}$, $\Delta$$\delta_{\delta}$) =
(-0.007 $\pm$ 0.032, -0.030 $\pm$ 0.030) mas, indicating that there is no evidence for a bias between
HST-Gaia and Gaia-DR3 positions to a total precision of
$\sigma_{\Delta pos}$ = $\sqrt{\sigma_{\Delta \delta_{\alpha^*}}^2 + \sigma_{\Delta \delta_{\delta}}^2}$ = 0.044 mas.

\subsection{Transforming the HST-Gaia Proper Motions Into SgrA*-at-Rest Coordinates}
\label{sec:sgra_rest}
We convert the HST-Gaia measurements from ICRS coordinates into SgrA*-at-Rest coordinates
using the ICRS position and proper motion of SgrA* measured
by \citet{Xu:2022sx}\footnote{\citet{Xu:2022sx} measure the position of SgrA* to be $\alpha$(J2000) = 17$^h$45$^m$40$^s$.032863 $\pm$ 0$^s$.000016,
$\delta$(J2000) = $-$29$^{\circ}$00$'$28.$''$24260 $\pm$ 0$''$.00047 at a reference epoch of 2020.0 with a proper motion of
$ \mu_{\alpha^*, sgra}$ = -3.152 $\pm$ 0.011 mas yr$^{-1}$,  $\mu_{\delta, sgra}$ = -5.586 $\pm$ 0.006 mas yr$^{-1}$.}.
The SgrA*-at-Rest proper motions ($\mu_{\alpha*}^s$, $\mu_\delta^s$) are calculated
by subtracting the SgrA* ICRS proper motion from the HST-Gaia proper motions:

\begin{equation}
\mu_{\alpha*}^s = \mu_{\alpha*} - \mu_{\alpha*, sgra}
\end{equation}
\begin{equation}
\mu_{\delta}^s = \mu_{\delta} - \mu_{\delta, sgra}
\end{equation}

where $\mu_{\alpha*, sgra}$, $\mu_{\delta, sgra}$ is the SgrA* ICRS proper motion from \citet{Xu:2022sx}.
Calculating the SgrA*-at-Rest positions requires multiple steps since the ICRS position of SgrA* changes with time.
For each star, we first use the \citet{Xu:2022sx} proper motion to calculate the ICRS position
of SgrA* at the reference epoch of that star's HST-Gaia proper motion fit (i.e., $t_0$ in Equation \ref{eq:pm}).
We denote this position as ($\alpha_{sgra}$($t_0$), $\delta_{sgra}$($t_0$)).
Next, we calculate the offset between this position and the fiducial position of
the HST-Gaia measurements:

\begin{equation}
d\alpha* = (\alpha_{sgra}(t_0) - \alpha_{f}) * cos(\delta_{sgra})
\end{equation}
\begin{equation}
d\delta = \delta_{sgra}(t_0) - \delta_{f}
\end{equation}

Then, these offsets are subtracted from the reference epoch position of the HST-Gaia proper motion fit
to get the star's position relative to SgrA* at the reference epoch, denoted $\delta_{\alpha^*_0}^{s}$ and $\delta_{\delta_0}^{s}$:

\begin{equation}
\delta_{\alpha^*_0}^{s} = \delta_{\alpha*_0} - d\alpha*
\end{equation}
\begin{equation}
\delta_{\delta_0}^{s} = \delta_{\delta_0} - d\delta
\end{equation}

We justify adopting the SgrA* ICRS measurements from \citet{Xu:2022sx} over others in the literature \citep[e.g.][]{Reid:2020jo, Gordon:2023ck}
in Appendix \ref{app:sgra_motion}.

\section{Results}
\label{sec:results}

\subsection{A Catalog of HST-Gaia Stars Within the Central Parsec of the Galaxy}
\label{sec:ref_star_cat}
We present a proper motion catalog of 2876 stars derived from the
HST-Gaia measurements which represent astrometric
reference stars for the HST-Gaia reference frame within R $\leq$ 25" (1 pc) of SgrA*.
The process used to identify these stars is described in $\mathsection$\ref{sec:ref_star_select}.
The sample can be separated into 3 groups.
The first group is the 40 Gaia reference stars
that are used to transform the HST astrometry into the Gaia-CRF3 reference frame.
We refer to these as the primary reference stars since they define the HST-Gaia
coordinate system.
The second group contains 2823 well-measured stars within R $\lesssim$ 25"
that fall within footprint of ground-based AO imaging of the GC \citep[e.g.][]{Plewa:2015ud, Sakai:2019fm}.
These are the secondary reference stars which establish the HST-Gaia reference frame
in this region.
The third group is 13 stellar masers which overlap with the radio sample
used to define the most recent maser-based reference frame at the GC \citepalias{Darling:2023ao}.
These are used to assess the consistency between the HST-Gaia and radio maser measurements.
Note that many of the masers and primary reference stars are located beyond R $\geq$ 25''
from SgrA* but are included in this sample due to their relevance to the HST-Gaia reference frame.
The HST-Gaia measurements in ICRS coordinates
are reported in Table \ref{tab:hst_pm} along with their corresponding HST photometry
in Table \ref{tab:phot}.
In addition, the HST-Gaia measurements transformed into SgrA*-at-Rest coordinates
are provided in Appendix \ref{app:sgra_pm} for convenience.

\begin{figure*}
\begin{center}
\includegraphics[scale=0.3]{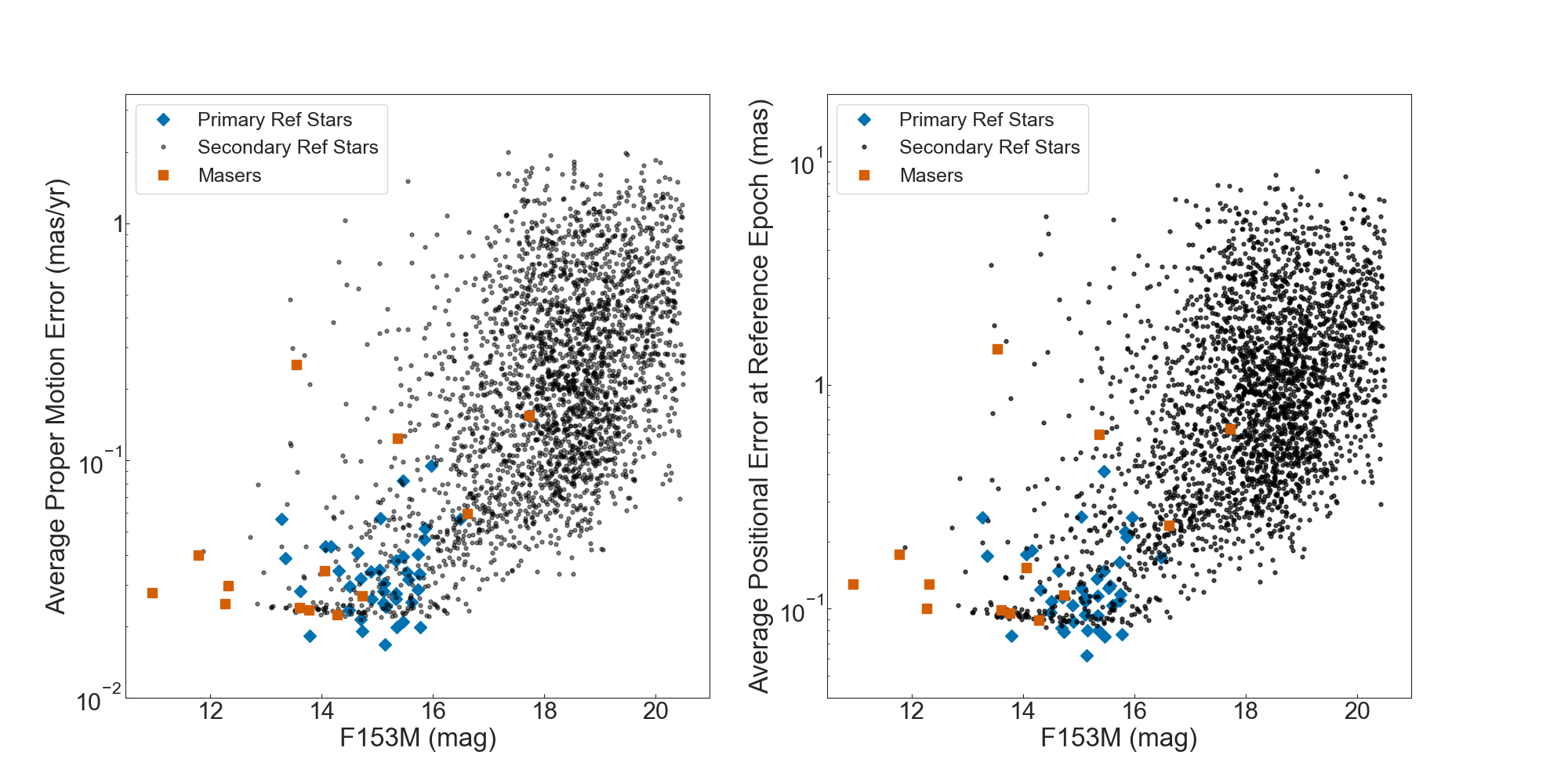}
\caption{Average HST-Gaia proper motion errors (left) and reference epoch position errors (right) as a function of F153M magnitude for the sample of stars presented in this paper.
Primary reference stars (from Gaia-DR3) are shown as blue diamonds, newly-created secondary reference stars within R $\lesssim$ 1pc from SgrA* as black points, and the masers
used for comparison with radio measurements as red squares. The data points represent the average of the $\alpha^*$ and $\delta$ errors for each star.
}
\label{fig:pm_errs}
\end{center}
\end{figure*}

The stars in this catalog span an HST magnitude range between 10.95 mag $<$ F153M $<$ 20.5 mag,
and the brightest sources (F153M $\leq$ 15 mag) achieve median
position and proper motion errors of
0.11 mas and  0.03 mas yr$^{-1}$, respectively (Figure \ref{fig:pm_errs}).
These measurements are $\sim$20x more precise than the
ICRS proper motions derived from the VVV survey field that covers
the GC region \citep{Griggio:2024dt}.
This improvement is
enabled by the astrometric performance of the HST WFC3-IR observations
and the time baseline of the dataset.

A majority of the stars in the catalog prefer a best-fit kinematic model with multiple kernels.
The most commonly preferred model is one that contains
a time-based correlation kernel:
47.6\% of the sample prefers a \emph{poly+sqexp} model and a further 17.2\% prefer
a \emph{poly+step} model.
Only a small number of stars prefer the
\emph{poly+confusion} (2.6\%) or \emph{poly+add} models (1.4\%).
The remaining 31.2\% of the sample is best described using a \emph{poly-only}
model, indicating that their astrometry can be described by linear
proper motion alone.
The fraction of stars that prefer a \emph{poly-only} model is higher
for brighter stars ($\sim$54\% of stars with F153M $<$ 16 mag)
compared to fainter stars ($\sim$28\% of stars with F153M $>$ 16 mag).
This distribution suggests that the systematic errors in the
HST astrometry are dominated by the PA-dependent shifts
discussed in Appendix \ref{app:ast_errs}, which manifest as
a correlation in time and is stronger for fainter stars than for brighter ones.

\begin{rotatetable}
\movetableright=3mm
\begin{deluxetable*}{lccccccccccccccc}
\tablewidth{0pt}
\tabletypesize{\tiny}
\tablecaption{HST-Gaia Proper Motions: ICRS Coordinates}
\tablehead{
\colhead{Name} & \colhead{$\alpha$} & \colhead{$\delta$} & \colhead{$\sigma_{\alpha*}$}& \colhead{$\sigma_{\delta}$} &
\colhead{$\mu_{\alpha*}$} & \colhead{$\sigma_{\mu_{\alpha*}}$} & \colhead{$\mu_{\delta}$} & \colhead{$\sigma_{\mu_{\delta}}$} & \colhead{$t0$} & \colhead{$\widetilde{\chi^2_{\mathrm{loo}}}({\alpha*})$} & \colhead{$\widetilde{\chi^2_{\mathrm{loo}}}({\delta}$)}  & \colhead{$N$} & \colhead{Model} & \colhead{Ref} &\colhead{Alt Name}\\
& (deg) & (deg) & (mas) & (mas) & (mas/yr) & (mas/yr) & (mas/yr) & (mas/yr) & (year) &  &  &  & &  &
}
\startdata
HST\_NSC\_007314 & 266.41680951 & -29.00792399 & 3.76 & 3.84 & -8.97 & 0.90 & 4.47 & 0.89 & 2018.4467 & 1.08 & 0.99 & 11 & 1 & 0 & S0-6 \\
HST\_NSC\_007079 & 266.41689152 & -29.00800914 & 1.01 & 1.00 & 4.86 & 0.25 & -10.73 & 0.25 & 2015.4906 & 1.75 & 2.00 & 9 & 0 & 0 & S0-9 \\
HST\_NSC\_006757 & 266.41697611 & -29.00795541 & 0.43 & 0.42 & 0.16 & 0.18 & -6.44 & 0.18 & 2019.6907 & 0.76 & 0.56 & 14 & 0 & 0 & S0-13 \\
HST\_NSC\_007316 & 266.41663396 & -29.00772225 & 5.06 & 1.06 & -2.26 & 0.94 & -3.11 & 0.30 & 2019.3942 & 1.16 & 0.59 & 12 & 2 & 0 & S0-12 \\
HST\_NSC\_007214 & 266.41657304 & -29.00792728 & 0.40 & 0.46 & -0.20 & 0.08 & -7.48 & 0.10 & 2018.5698 & 0.90 & 0.96 & 14 & 2 & 0 & S0-14 \\
HST\_NSC\_007322 & 266.41690889 & -29.00759950 & 1.25 & 1.25 & -11.58 & 0.38 & -2.59 & 0.38 & 2012.8189 & 1.27 & 1.80 & 9 & 0 & 0 & S1-3 \\
HST\_NSC\_006751 & 266.41690777 & -29.00807787 & 5.72 & 5.27 & -6.56 & 0.92 & -1.09 & 0.88 & 2013.5218 & 0.43 & 0.86 & 11 & 2 & 0 & S1-5 \\
HST\_NSC\_016387 & 266.41654287 & -29.00772010 & 9.01 & 2.66 & 4.42 & 1.83 & -7.26 & 0.66 & 2017.8891 & 0.60 & 1.03 & 9 & 5 & 0 & S1-26 \\
HST\_NSC\_007323 & 266.41715087 & -29.00782874 & 0.84 & 0.58 & 2.01 & 0.16 & -4.64 & 0.13 & 2017.6056 & 0.65 & 0.90 & 14 & 2 & 0 & S1-1 \\
HST\_NSC\_006701 & 266.41711206 & -29.00767334 & 0.24 & 0.22 & -12.59 & 0.05 & 1.45 & 0.04 & 2018.6155 & 0.60 & 0.66 & 13 & 2 & 0 & irs16C \\
\enddata
\tablecomments{Description of columns: \emph{Name}: HST star name,
\emph{$\alpha$, $\delta$}: RA and DEC position at $t0$,
\emph{$\sigma_{\alpha*}$, $\sigma_{\delta}$}: Uncertainty in $\alpha*$ and $\delta$, where $\sigma_{\alpha*} = \sigma_{\alpha} * cos(\delta)$,
\emph{$\mu_{\alpha*}$, $\mu_{\delta}$}: HST-Gaia proper motions,
\emph{$\sigma_{\mu_{\alpha*}}$, $\sigma_{\mu_{\delta}}$}: error in $\mu_{\alpha*}$ and $\mu_{\delta}$,
\emph{$t0$}: reference epoch of the HST-Gaia proper motions (calculated as the astrometric-error weighted average time across the epoch dates in Table \ref{tab:obs}),
\emph{$\widetilde{\chi^2_{\mathrm{loo}}}({\alpha*})$, $\widetilde{\chi^2_{\mathrm{loo}}}({\delta})$}: LOO Reduced $\chi^2$ values for $\mu_{\alpha*}$ and $\mu_{\delta}$ fits, respectively (Eqn \ref{eq:chi2_red}),
\emph{$N$}: Number of F153M epochs observed,
\emph{Model}: Specifies the kinematic model used -- 0: \emph{poly-only}, 1: \emph{poly+conf}, 2: \emph{poly+sqexp}, 3: \emph{poly+add}, 5: \emph{poly+step},
\emph{Ref}: Type of reference star -- 0: Secondary reference star, 1: primary reference star, 2: maser,
\emph{Alt Name}: Alternative name for star, if applicable (either from \citet{Sakai:2019fm} or the Gaia-DR3 catalog)
}
\tablecomments{The proper motion fit uncertainties reported in this table ($\sigma_{\alpha*}$, $\sigma_{\delta}$, $\sigma_{\mu_{\delta}}$, $\sigma_{\mu_{\delta}}$)
do not include the systematic error between the HST-Gaia and Gaia-CRF3 discussed in $\mathsection$\ref{sec:ref_frame_comp}.}
\tablecomments{The full table is available in machine-readable format.}
\label{tab:hst_pm}
\end{deluxetable*}
\end{rotatetable}

\begin{deluxetable*}{lcccccccc}
\tablewidth{0pt}
\tabletypesize{\tiny}
\tablecaption{HST Photometry}
\tablehead{
\colhead{Name} & \colhead{F127M} & \colhead{$\sigma_{F127M}$} & \colhead{F139M} & \colhead{$\sigma_{F139M}$} & \colhead{F153M} & \colhead{$\sigma_{F153M}$} & \colhead{Ref Type} & \colhead{Alt Name}
}
\startdata
HST\_NSC\_007314 & 18.73 & 0.03 & 17.40 & 0.06 & 16.07 & 0.05 & 0 & S0-6 \\
HST\_NSC\_007079 & 19.10 & 0.03 & 17.87 & 0.06 & 16.70 & 0.19 & 0 & S0-9 \\
HST\_NSC\_006757 & 18.38 & 0.02 & 17.07 & 0.02 & 15.70 & 0.13 & 0 & S0-13 \\
HST\_NSC\_007316 & 20.99 & 0.15 & 19.17 & 0.09 & 17.68 & 0.09 & 0 & S0-12 \\
HST\_NSC\_007214 & 18.85 & 0.04 & 17.64 & 0.03 & 16.42 & 0.08 & 0 & S0-14 \\
HST\_NSC\_007322 & -999.00 & -999.00 & -999.00 & -999.00 & 14.20 & 0.05 & 0 & S1-3 \\
HST\_NSC\_006751 & 18.78 & 0.02 & 17.31 & 0.03 & 15.62 & 0.12 & 0 & S1-5 \\
HST\_NSC\_016387 & 20.94 & 0.30 & 19.54 & 0.22 & 17.87 & 0.15 & 0 & S1-26 \\
HST\_NSC\_007323 & 18.01 & 0.02 & 16.77 & 0.02 & 15.59 & 0.02 & 0 & S1-1 \\
HST\_NSC\_006701 & 15.20 & 0.01 & 13.93 & 0.01 & 12.72 & 0.13 & 0 & irs16C \\
\enddata
\label{tab:phot}
\tablecomments{Description of columns: \emph{Name}: star name,
\emph{F127M, F139M, F153M}: Vega mags in corresponding filters (F153M is the average mag across all F153M epochs),
\emph{$\sigma_{F127M}$, $\sigma_{F139M}$, $\sigma_{F153M}$}: error in corresponding mags ($\sigma_{F153M}$ is the average error across all F153M epochs),
\emph{Ref Code}: Type of ref star -- 0: Secondary reference star, 1: primary reference star, 2: maser,
\emph{Alt Name}: Alternative name for star, if applicable \citep[e.g. from][or Gaia-DR3 catalog]{Sakai:2019fm}
}
\tablecomments{Values of -999.0 indicate that the star wasn't detected in that filter. The full table is available in machine-readable format.}
\end{deluxetable*}

\clearpage

\subsubsection{Identifying the HST-Gaia Astrometric Reference Stars}
\label{sec:ref_star_select}
The selection of the 40 primary reference stars is described in detail in $\mathsection$\ref{sec:gaia_ref}.
Here we discuss the identification of the secondary reference stars and masers.

The secondary reference stars comprise of well-measured HST-Gaia sources that
are located within the 22'' x 22'' mosaic of ground-based AO observations
described by \citet[][]{Sakai:2019fm}, which has been observed annually since 2005 as part of the Galactic Center Orbits Initiative
\citep[PI: A.M. Ghez; see also][]{Ghez:2008tg, Yelda:2010fu, Jia:2019oq}.
To remove sources with poor HST-Gaia astrometry,
we require that the secondary reference stars meet the following
criteria:
(1) they must be detected in at least 4 HST epochs
and have kinematic motion fits with $\widetilde{\chi^2_{\mathrm{loo}}} \leq$ 5 (Equation \ref{eq:chi2_red});
(2) have reference epoch position uncertainties~$<$~10~mas and proper motion
uncertainties~$<$~2 mas yr$^{-1}$;
(3) have reference epoch position and proper motion uncertainties  $>$0.01 mas and $>$0.01 mas yr$^{-1}$,
respectively, removing a small fraction of stars where the kinematic model fit fails and returns unphysically small errors ($\sim$1\% of total sample);
(3) have a total proper motion $<$ 15 mas/yr, to eliminate erroneous high-proper motion sources caused by
stellar mismatches or confusion;
and (4) have F153M $\leq$ 20.5 mag.
Of an initial set of 3250 HST-Gaia sources within the \citet[][]{Sakai:2019fm} field,
2823 fulfill these criteria and define the secondary reference star sample.

The final set of stars in this catalog are the NIR counterparts to the stellar masers
used to define the most recent radio-based astrometric reference frame
at the GC from \citetalias{Darling:2023ao}.
To be matched with a maser source, the HST-Gaia position must
match within 0.1$''$ at a common epoch of 2015.5.
Of the 15 masers in the \citetalias{Darling:2023ao} sample, 13 are successfully matched with HST-Gaia
sources.

\subsection{The Consistency Between the HST-Gaia and Radio Maser Reference Frames at the GC}
\label{sec:ref_frame}
The consistency between the the HST-Gaia reference frame and the
most recent radio-based maser reference frame at the GC \citepalias{Darling:2023ao} is evaluated by comparing
the positions and proper motions of the stellar masers present in both catalogs.
However, the radio measurements of the masers are made in a SgrA*-at-Rest coordinate system,
and so the HST-Gaia measurements must be transformed into the same system
as discussed in $\mathsection$\ref{sec:sgra_rest}.
Thus, the additional uncertainties between the Gaia-CRF3 and ICRS
coordinate system ($\mathsection$\ref{sec:hst_gaia_icrf3}) and
the SgrA* ICRS measurements themselves ($\mathsection$\ref{sec:icrf3_sgra}) must be taken into account.
These uncertainties are summarized in Table \ref{tab:ref_frame_errs} and
the final comparison between the HST-Gaia and \citetalias{Darling:2023ao}
radio measurements is presented in $\mathsection$\ref{sec:hst_gaia_masers}.

\subsubsection{Systematic Uncertainties Between HST-Gaia and ICRS}
\label{sec:hst_gaia_icrf3}
Since the orientation of Gaia-CRF3 is fixed to ICRS, the HST-Gaia reference frame
is also consistent with ICRS.
To determine the systematic uncertainty between HST-Gaia and ICRS,
the systematic uncertainty between Gaia-CRF3 and ICRS must be considered.

Gaia-CRF3 is aligned to the ICRS using quasar sources in the Gaia-DR3 catalog,
with $\sim$0.4 million sources used to fix the rotational state
of the reference frame (i.e., asserting that the proper motions of these
extragalactic sources are consistent with zero) and
$\sim$2000 sources with counterparts in the third realization of the
International Celestial Reference Frame \citep[ICRF3;][]{Charlot:2020tn}
to constrain the orientation \citep{Gaia-Collaboration:2022cm}.
Maps of the proper motions of over 1 million quasar sources in Gaia-DR3
reveal systematic errors of  0.0112 mas yr$^{-1}$ in $\mu_{\alpha^*}$ and 0.0107 mas yr$^{-1}$ in $\mu_{\delta}$ \citep{Lindegren:2021ae, Gaia-Collaboration:2022cm},
which we adopt as the systematic proper motion uncertainty between Gaia-CRF3 and ICRS.
The uncertainty between the Gaia-CRF3 and ICRS positions is estimated from
the position differences of common quasar sources between Gaia-DR3 and ICRF3.
\citet{Gaia-Collaboration:2022cm} find the median position difference ($\alpha^*$ and $\delta$ combined)
to be 0.516 mas across the 3142 such sources.
This suggests an error-on-the-mean difference (analogous to Equation \ref{eq:wEOM}) of 0.516 mas / $\sqrt{3142}$ = 0.009 mas.
Assuming this is evenly split between $\alpha^*$ and $\delta$ implies a systematic position uncertainty of
0.007 mas in each direction\footnote{\citet{Gaia-Collaboration:2022cm} find
the median position offsets across the quasar sample (analogous to Eqn \ref{eq:wmean}) to be -0.004 mas and 0.006 mas in $\alpha^*$ and $\delta$, respectively, which
is consistent with zero to within the estimated uncertainty of 0.007 mas in each direction.}.

The total systematic uncertainty between HST-Gaia and ICRS is calculated as
the quadratic sum of the systematic uncertainty between HST and Gaia-CRF3 ($\mathsection$\ref{sec:gaia_ref}) and
the systematic uncertainty between Gaia-CRF3 and ICRS.
The total systematic uncertainty in proper motion is thus
($\sigma_{\Delta \mu_{\alpha^*}}$, $\sigma_{\Delta \mu_{\delta}}$, $\sigma_{\Delta pm}$) = (0.023, 0.018, 0.029) mas/yr,
where $\sigma_{\Delta pm}$ = $\sqrt{\sigma_{\Delta \mu_{\alpha^*}}^2 + \sigma_{\Delta \mu_{\delta}}^2}$.
The total systematic uncertainty in position is
($\sigma_{\Delta \alpha^*}$, $\sigma_{\Delta \delta}$, $\sigma_{\Delta pos}$) = (0.033, 0.031, 0.045) mas,
where $\sigma_{\Delta pos} = \sqrt{\sigma_{\Delta \alpha^*}^2 + \sigma_{\Delta \delta}}$.
These uncertainties are overwhelmingly dominated by the uncertainty between HST and Gaia-CRF3,
with the Gaia-CRF3 to ICRS step making only a very small contribution.

\subsubsection{HST-Gaia and SgrA*-at-Rest}
\label{sec:icrf3_sgra}
Transforming HST-Gaia measurements into a SgrA*-at-Rest frame
incurs the additional uncertainty in the
ICRS measurements of SgrA* \citep{Xu:2022sx}.
These uncertainties are
($\sigma_{\mu_{\alpha^*}}$, $\sigma_{\mu_{\delta}}$) = (0.011, 0.006) mas/yr
and ($\sigma_{\alpha^*}$, $\sigma_{\delta}$) = (0.21, 0.47) mas.
The total uncertainty between HST-Gaia and the SgrA*-at-Rest frame
is then the quadratic sum of the uncertainties between HST-Gaia and Gaia-CRF3,
Gaia-CRF3 and ICRS, and the SgrA* measurements.
This yields ($\sigma_{\Delta \mu_{\alpha^*}}$, $\sigma_{\Delta \mu_{\delta}}$, $\sigma_{\Delta pm}$) $\sim$ (0.025, 0.019, 0.031) mas/yr
and ($\sigma_{\Delta \alpha^*}$, $\sigma_{\Delta \delta}$, $\sigma_{\Delta pos}$) $\sim$ (0.213, 0.471, 0.517) mas.
The total proper motion uncertainty is dominated by the uncertainty between the
HST-Gaia and Gaia-CRF3 frames while the total position uncertainty
is dominated by the ICRS position of SgrA*.

\subsubsection{Constraining the Bias Between HST-Gaia and Radio Maser Reference Frames}
\label{sec:hst_gaia_masers}
We compare the positions and proper motions of 13 masers from \citetalias{Darling:2023ao} that
are successfully matched in the HST-Gaia sample\footnote{The 13 masers used to compare the HST-Gaia and \citetalias{Darling:2023ao} radio maser reference frames are
IRS7, IRS12N, IRS9, IRS10EE, IRS15NE, IRS28, IRS17, SiO-15, IRS19NW,
SiO-14, SiO-11, SiO-16, and SiO-6.}.
The extent to which a bias may be present between the two frames is quantified
using Equations \ref{eq:wmean} and \ref{eq:wEOM}.
To measure the position bias, the HST-Gaia and \citetalias{Darling:2023ao} proper
motions are used to calculate the maser positions at a common epoch
of 2017.4891 (the optimal comparison epoch as calculated via Equation \ref{eq:t0_comp}).

\begin{figure*}
\begin{center}
\includegraphics[scale=0.3]{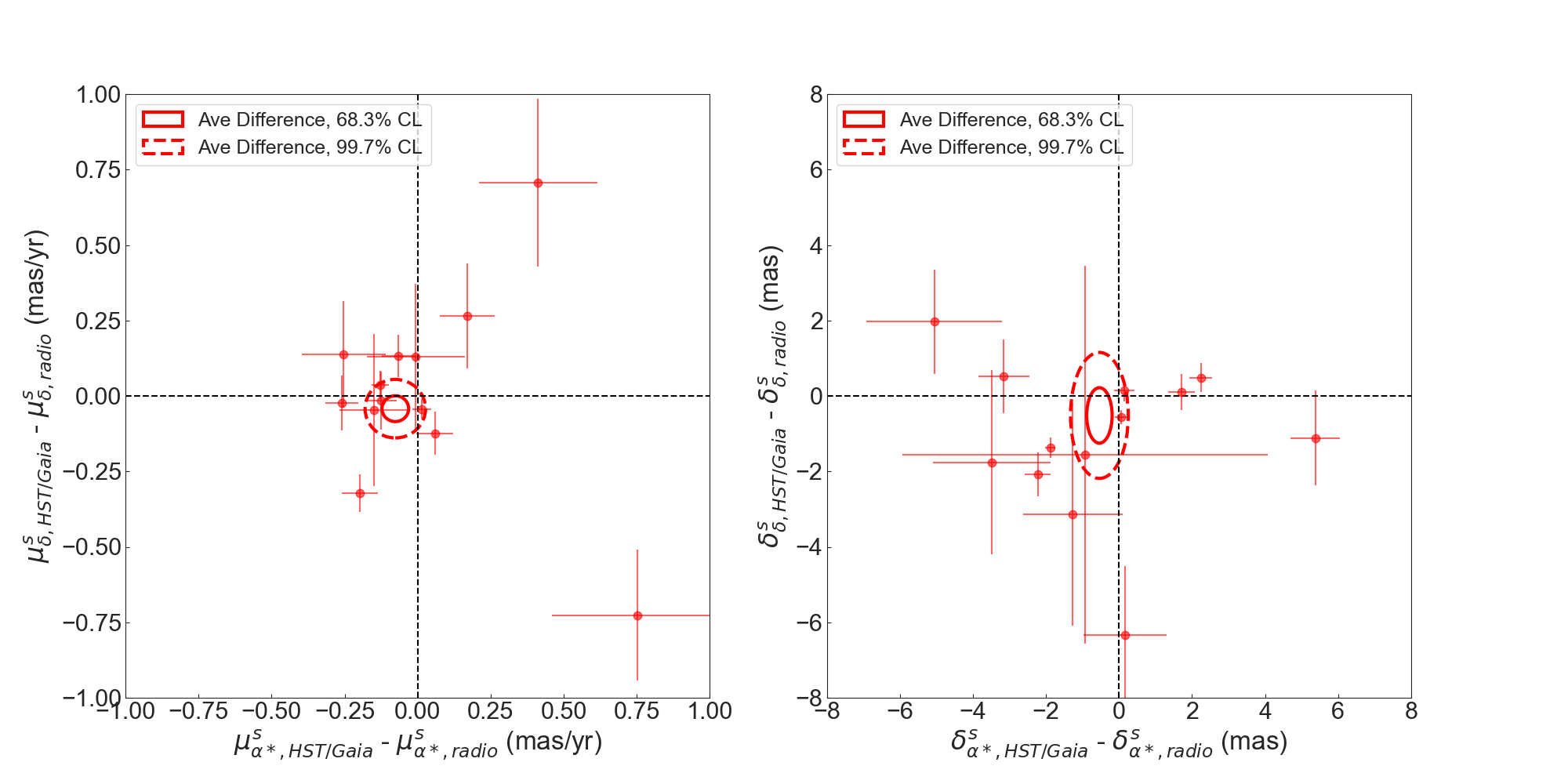}
\caption{Proper motion (left) and position (right) differences for the masers between the HST-Gaia and
radio measurements in SgrA*-at-Rest coordinates (red points).
We calculate an average offset of (-0.076 $\pm$ 0.030, -0.042 $\pm$ 0.028) mas yr$^{-1}$
in the ($\mu_{\alpha*}^s$, $\mu_{\delta}^s$) proper motions and (-0.536 $\pm$ 0.231, -0.517 $\pm$ 0.486) mas in the ($\delta_{\alpha*}^s$, $\delta_{\delta}^s$) positions (at a common epoch of 2017.4891).
The red circle represents the 68.3\% confidence level and the red dashed circle represents the 99.7\% confidence level for the average offset.
This indicates that the HST-Gaia and radio reference frames are consistent (i.e., the average offset is consistent with zero) to within 99.7\% confidence,
although with a possible bias in the $\alpha*$ directions (in both position and proper motion)
at $\sim$2.5$\sigma$ significance.
}
\label{fig:maser_diffs}
\end{center}
\end{figure*}

From the maser measurements alone
(i.e. ignoring the systematic reference frame uncertainties from Table \ref{tab:ref_frame_errs}),
we find the average proper motion differences to be
(-0.076 $\pm$ 0.016, -0.042 $\pm$ 0.020) mas/yr in ($\mu_{\alpha*}^s$, $\mu_{\delta}^s$)
and (-0.536 $\pm$ 0.088, -0.517 $\pm$ 0.119) mas in ($\delta_{\alpha*}^s$, $\delta_{\delta}^s$).
When the systematic reference frame uncertainties are
added in quadrature to the measurement uncertainties, the differences between the HST-Gaia and radio maser frames become
(-0.076 $\pm$ 0.030, -0.042 $\pm$ 0.028) mas/yr in ($\mu_{\alpha*}^s$, $\mu_{\delta}^s$)
and (-0.536 $\pm$ 0.230, -0.517 $\pm$ 0.486) mas in ($\delta_{\alpha*}^s$, $\delta_{\delta}^s$)  (Figure \ref{fig:maser_diffs}).
Thus, the joint probability distribution of the differences in the ($\mu_{\alpha*}^s$, $\mu_{\delta}^s$) proper motions
are consistent with zero to within 0.041 mas yr$^{-1}$ at 99.7\% confidence.
Similarly, the joint probability distribution of the differences in the ($\delta_{\alpha*}^s$, $\delta_{\delta}^s$) positions
are consistent with zero to within 0.54 mas at  99.7\% confidence.

However, a possible tension at may be present in the $\alpha*$ direction in
both proper motion and position
at $\sim$2.5$\sigma$ significance, which can be seen
by the apparent leftward shift in the average differences from zero in Figure \ref{fig:maser_diffs}.
This tension could indicate the presence of a systematic error in either the
HST-Gaia reference frame, radio maser reference frame, the measured ICRS
position and proper motion of SgrA* from \citet{Xu:2022sx}, or some combination
of the three.
Further improvements in the HST-Gaia and radio measurements are required
to determine if this tension is evidence for a true offset or not.

\begin{deluxetable*}{l l l l l l l l l}
\tabletypesize{\footnotesize}
\label{tab:ref_frame_errs}
\tablecaption{The Systematic Uncertainties Between Reference Frames}
\tablehead{
\colhead{Reference Frame} & Coordinate System & \colhead{$\sigma_{\Delta \mu_{\alpha^*}}$} & $\sigma_{\Delta \mu_{\delta}}$ & \colhead{$\sigma_{\Delta pm}$} & \colhead{$\sigma_{\Delta \alpha^*}$} & \colhead{$\sigma_{\Delta \delta}$} & $\sigma_{\Delta pos}$ & \colhead{Notes} \\
& & (mas yr$^{-1}$) & (mas yr$^{-1}$) & (mas yr$^{-1}$) & (mas) & (mas) & (mas) &
}
\startdata
HST-Gaia & Gaia-CRF3 & 0.02 & 0.015 & 0.025 & 0.032 & 0.03 & 0.044 & see $\mathsection$\ref{sec:gaia_ref} \\
HST-Gaia & ICRS & 0.023 & 0.018 & 0.029 & 0.033 & 0.031 & 0.045 & see $\mathsection$\ref{sec:hst_gaia_icrf3}\\
HST-Gaia & SgrA*-at-Rest & 0.025 &  0.019 & 0.031 & 0.213 & 0.471 & 0.517 & see $\mathsection$\ref{sec:icrf3_sgra} \\
\citet{Plewa:2015ud} & SgrA*-at-Rest & 0.08 & 0.07 & 0.11 & 0.17 & 0.17 & 0.24 & AO obs of masers \\
\citet{Sakai:2019fm} & SgrA*-at-Rest & 0.018 & 0.025 & 0.031 & 0.458 & 0.455 & 0.646 & AO obs of masers \\
\enddata
\tablecomments{As discussed in $\mathsection$\ref{sec:hst_gaia_icrf3},  $\sigma_{\Delta pm}$ = $\sqrt{\sigma_{\Delta \mu_{\alpha^*}}^2 + \sigma_{\Delta \mu_{\delta}}^2}$ and $\sigma_{\Delta pos} = \sqrt{\sigma_{\Delta \alpha^*}^2 + \sigma_{\Delta \delta}}$.}
\end{deluxetable*}

\section{Discussion}
\label{sec:discussion}

\subsection{Uncertainties in the HST-Gaia Measurements and Reference Frame: Comparison with Past Work}
\label{sec:ref_comp_section}
The GC reference frame is a key source of uncertainty
when studying the motions and orbits of stars near SgrA* \citep[e.g.][]{Do:2019gr, The-GRAVITY-Collaboration:2024xt}.
Here we compare two facets of the uncertainty in the HST-Gaia reference
frame to previous GC NIR reference frames in the literature:
the overall precision of the reference frame (i.e., the uncertainty in the systematic bias relative to
the desired coordinate system)
and the uncertainties in the individual position and proper motion measurements for the astrometric
reference stars.
In summary, the HST-Gaia reference frame offers an improved
overall uncertainty compared to
past GC NIR reference frames constructed using AO observations of stellar masers,
thus lowering the error floor that can be achieved ($\mathsection$\ref{sec:overall_ref_error}).
However, the individual measurements uncertainties of the astrometric reference stars are
generally higher in HST-Gaia compared to AO maser-based frames,
which can lead to larger transformation errors when transforming astrometry into
the reference frame ($\mathsection$\ref{sec:ref_star_errs}).

\subsubsection{The Overall Precision of the Reference Frame}
\label{sec:overall_ref_error}
As discussed in $\mathsection$\ref{sec:ref_frame_comp}, the overall precision of a reference
frame is quantified by the uncertainty in the systematic bias
of the reference frame relative to its desired coordinate system.
This represents the maximum possible precision that can be achieved
for individual measurements in that reference frame.
The uncertainty in the systematic bias of the HST-Gaia reference frame (relative to Gaia-CRF3) is
($\sigma_{\Delta \mu_{\alpha^*}}$, $\sigma_{\Delta \mu_{\delta}}$, $\sigma_{\Delta pm}$) = (0.020, 0.015, 0.025) mas yr$^{-1}$
in proper motion and
($\sigma_{\Delta \alpha^*}$, $\sigma_{\Delta \delta}$, $\sigma_{\Delta pos}$) = (0.032, 0.030, 0.044) mas in position,
which compares favorably to previous GC NIR reference frames constructed via ground-based AO observations
of stellar masers (Table \ref{tab:ref_frame_errs}).
In terms of the proper motion, HST-Gaia
is $\sim$4x more precise than the reference frame from \citet{Plewa:2015ud},
which achieves ($\sigma_{\Delta \mu_{\alpha^*}}$, $\sigma_{\Delta \mu_{\delta}}$, $\sigma_{\Delta pm}$) = (0.08, 0.07, 0.11) mas yr$^{-1}$,
and is $\sim$1.2x more precise than the reference frame from \citet{Sakai:2019fm},
which achieves ($\sigma_{\Delta \mu_{\alpha^*}}$, $\sigma_{\Delta \mu_{\delta}}$, $\sigma_{\Delta pm}$) = (0.018, 0.025, 0.031) mas yr$^{-1}$.
In terms of the position, the HST-Gaia reference frame
is $\sim$5x more precise than \citet{Plewa:2015ud},
which achieves ($\sigma_{\Delta \alpha^*}$, $\sigma_{\Delta \delta}$, $\sigma_{\Delta pos}$) = (0.17, 0.17, 0.24) mas,
and is $\sim$15x more precise than \citet{Sakai:2019fm},
which achieves ($\sigma_{\Delta \alpha^*}$, $\sigma_{\Delta \delta}$, $\sigma_{\Delta pos}$) = (0.458, 0.455, 0.646) mas\footnote{The large position
error in \citet{Sakai:2019fm} is primarily due the Keck NIRC2 distortion correction error of 1 mas added in quadrature to the astrometric errors.}.
A key advantage of the HST-Gaia reference frame for reducing this error term is the
larger number of primary reference stars available to define it (40 Gaia-DR3 sources,
compared to only 7 or 8 masers for the AO maser-based frames).

When studying the orbit of the star S0-2/S2, recent studies have incorporated
additional constraints into the GC reference frame in order to
achieve even higher precision.
Such constraints include NIR position measurements of SgrA* when it is
in a bright state \citep[e.g.][]{GRAVITY-Collaboration:2020ro}
as well as constraints on the joint dynamical center of multiple Keplarian orbits
measured for stars in the region \citep{GRAVITY-Collaboration:2022kk, The-GRAVITY-Collaboration:2024xt}.
These studies report reference frame uncertainties of $\sigma_{\Delta pm}$ $\lesssim$ 0.01 mas yr$^{-1}$
and $\sigma_{\Delta pos}$ $\lesssim$ 0.01 mas. However, in many cases
the position and/or proper motion of SgrA* inferred by the stellar orbits
are not consistent with zero within these errors as would be expected for the
coordinate system.
A similar approach of combining the HST-Gaia reference frame with these
additional constraints is beyond the scope of this paper.

\subsubsection{Position and Proper Motion Uncertainties of Astrometric Reference Stars in the Central Parsec}
\label{sec:ref_star_errs}
The precision to which individual astrometric measurements can be transformed into
a given reference frame (i.e., the transformation errors) is related to the position and proper motion
uncertainties of the astrometric reference stars.
The lower the uncertainties of the reference stars, then the lower the transformation errors will be.
Here we compare the uncertainties of the HST-Gaia secondary reference
stars to the uncertainties of the secondary astrometric standards of
the AO maser-based reference frame at the GC from \citet{Sakai:2019fm}.

\begin{figure*}
\includegraphics[scale=0.35]{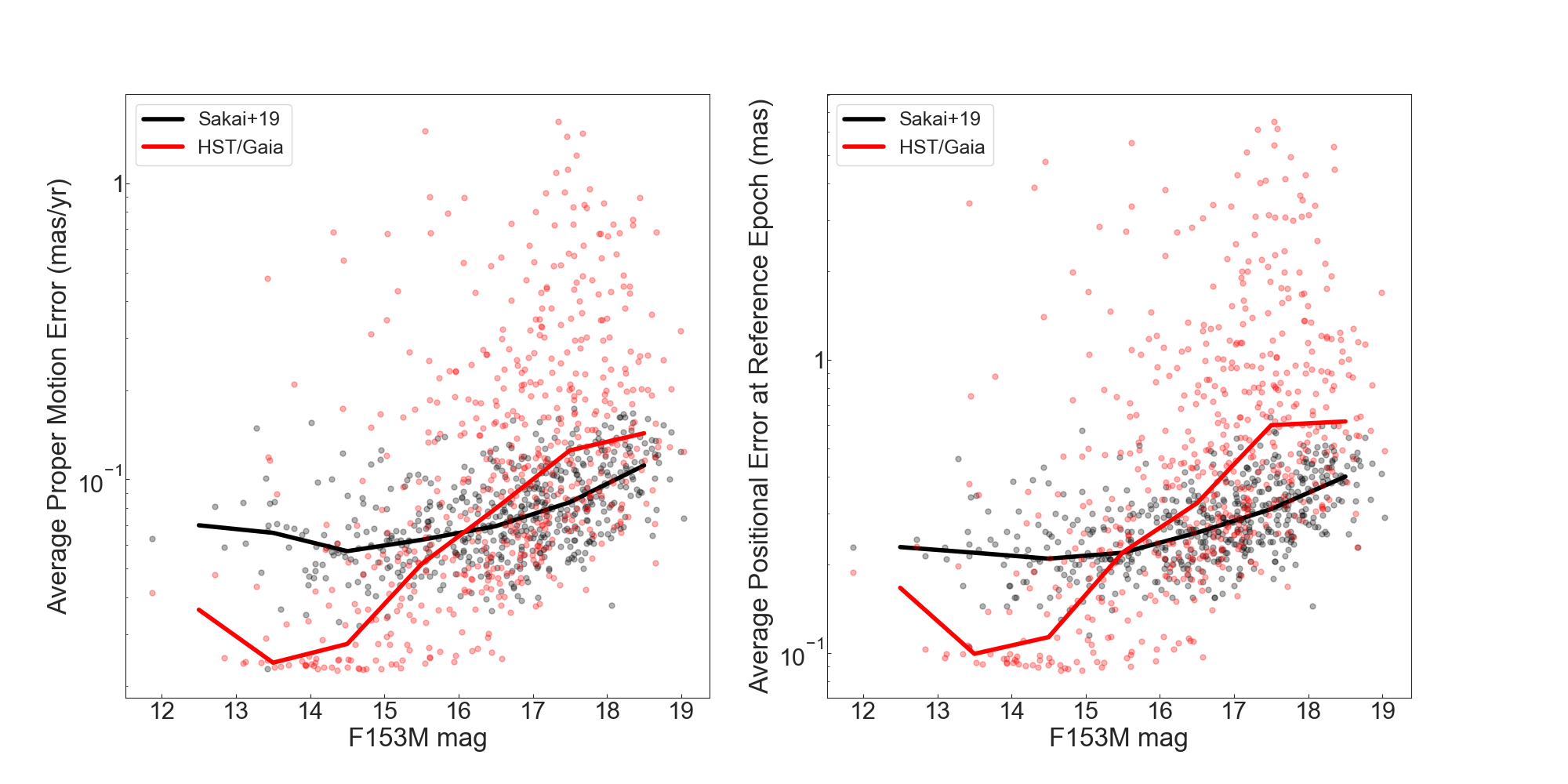}
\caption{The average proper motions errors (left) and reference epoch position errors (right) as a function
of F153M magnitude for the 542 stars in common between the HST-Gaia and \citet{Sakai:2019fm} reference star
catalogs. The HST-Gaia measurements are shown by the red points and the AO measurements by the black points,
with the corresponding lines representing the median errors as a function of magnitude.
The precision of the HST-Gaia measurements are comparable to the AO
measurements for bright sources (F153M $\leq$ 16 mag) while the AO measurements are more precise
for fainter sources (F153M $>$ 16 mag).}
\label{fig:sec_err_comp}
\end{figure*}

A total of 542 stars are found in common between the HST-Gaia catalog and \citet{Sakai:2019fm} reference star catalogs.
The AO measurements are more precise for most sources, with $\sim$68\% of reference stars
having more precise AO positions and $\sim$62\% of reference stars having more precise AO
proper motions when compared to the corresponding HST-Gaia measurements (Figure \ref{fig:sec_err_comp}).
This is not unexpected given that the AO observations
have higher spatial resolution than HST WFC3-IR (FWHM $\sim$ 60 mas vs. FWHM $\sim$ 170 mas)
and are thus less impacted by stellar crowding.
However, for bright sources (F153M $\leq$ 16 mag), the HST-Gaia measurements are more comparable
to the AO measurements,
with similar median
proper motion errors (0.05 mas yr$^{-1}$ and 0.06 mas yr$^{-1}$, respectively) and
median position errors (0.20 mas and 0.22 mas, respectively).
Stellar confusion has less impact on the astrometry of the brighter sources and
so the lower-resolution HST-Gaia measurements become more competitive.
On the other hand, the impact of stellar confusion on the HST-Gaia measurements becomes
apparent for fainter sources (F153M $>$ 16 mag), where the AO
measurements have lower median proper motion errors (0.08 mas yr$^{-1}$
vs. 0.13 mas yr$^{-1}$) and position errors (0.30 mas vs. 0.55 mas).
The HST-Gaia measurement errors also exhibit significantly increased scatter in this regime (Figure \ref{fig:sec_err_comp}).

This comparison suggests that astrometry transformed into the HST-Gaia reference frame
may suffer from larger transformation errors compared to when an AO maser-based
reference frame is used.
One path to mitigate this is to restrict
the HST-Gaia reference star sample to the lower-error sources when
applying to data.
The optimal thresholds to adopt will change depending
on the requirements of the given science case as well as
the trade-off between using more HST-Gaia reference stars
with higher individual uncertainties vs. fewer HST-Gaia reference stars
with lower individual uncertainties.

\subsection{Expected Improvement in the HST-Gaia Reference Frame with Gaia-DR4}
The Gaia mission is planning to release a DR4 astrometric catalog sometime
after mid-2026 which is expected to deliver significantly improved astrometry
relative to the DR3 catalog\footnote{The Gaia mission release
schedule can be found at https://www.cosmos.esa.int/web/gaia/release and the predictions
for the astrometric performance of Gaia DR4 can be found at https://www.cosmos.esa.int/web/gaia/release.}.
These improvements will reduce the astrometric uncertainties of the Gaia primary reference stars
and thus also the transformation errors of the HST astrometry.
Here we quantify the expected improvement in the HST-Gaia astrometry and predict how it will
propagate into the HST-Gaia reference frame.

For a given HST epoch, the transformation uncertainty into the Gaia-CRF3 reference frame can be estimated as:

\begin{equation}
\label{eq:trans_err}
\sigma_{trans} = \alpha_{obs} \frac{\overline{\sigma_{pos}}}{\sqrt{N_{ref} - n_{params}}}
\end{equation}

where $\overline{\sigma_{pos}}$ is the average total position uncertainty of the
primary reference stars in that epoch,
$N_{ref}$ is the number of reference stars, $n_{params}$ is the number of parameters in the transformation
(6 for a second-order polynomial), and $\alpha_{obs}$ is an empirically-determined constant.
For each reference star, $\sigma_{pos}$ is the quadratic sum of the
HST measurement error and the Gaia catalog position error when propagated
to the HST epoch.
Using the current values from the transformations described in $\mathsection$\ref{sec:gaia_ref},
we find that $\alpha_{obs}$ $\approx$ 2.4.

Gaia DR4 is expected to reduce the Gaia reference epoch position and
proper motion uncertainties of the primary reference stars by an average factor of
$\sim$1.6x and $\sim$2.2x, respectively.
Assuming that no new reference stars are added\footnote{This is a conservative
assumption, as it is likely that the Gaia DR4 catalog will provide additional
primary reference stars in the HST field.} and recalculating $\overline{\sigma_{pos}}$ using these
improved Gaia uncertainties (while keeping the HST uncertainties the same),
we predict that $\sigma_{trans}$ will range between 0.14 mas and 0.30 mas
across the HST epochs.
This represents an average of $\sim$1.7x improvement over the current values
(e.g., Figure \ref{fig:trans_err}).

The constraints on the systematic bias between the HST-Gaia and Gaia-CRF3
reference frame is related to the proper motion precision
of the primary reference stars in both the HST-Gaia and Gaia catalog
measurements (Equation \ref{eq:wEOM}).
For the HST-Gaia measurements, the proper motion uncertainty of a given star goes
as approximately $\sigma_{pm}$ $\propto$ $\overline{\sigma_{ast}}$,
where $\overline{\sigma_{ast}}$ is the average astrometric error across the epochs.
The expected improvement in $\sigma_{trans}$ due to Gaia DR4 results in
an improvement of $\sim$1.4x in $\sigma_{ast}$ and thus the proper motion uncertainties
are expected to improve by the same factor.
Incorporating this improvement to the HST-Gaia measurements
and combining it with the improved Gaia DR4 measurements results in
a predicted uncertainty of
($\sigma_{\Delta \mu_{\alpha^*}}$, $\sigma_{\Delta \mu_{\delta}}$, $\sigma_{\Delta pm}$) $\approx$ (0.009, 0.009, 0.012) mas yr$^{-1}$
in the systematic bias between the HST-Gaia and Gaia-CRF3 reference frames.
Overall, this represents a factor of $\sim$2x improvement relative to current values.

\subsection{Applications of the HST-Gaia Reference Frame To GC Science}
\label{sec:applications}
\begin{figure}
\includegraphics[scale=0.35]{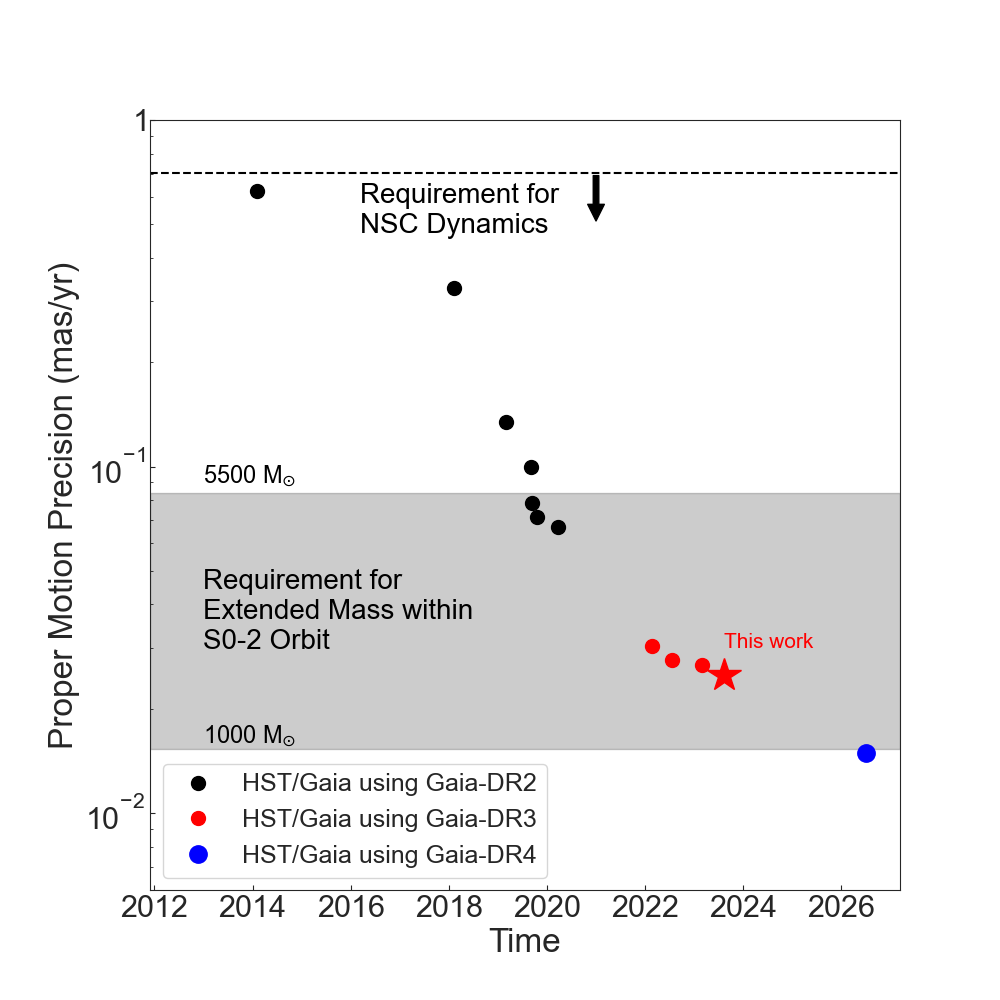}
\caption{The improvement in the proper motion precision of the HST-Gaia reference frame as a function of time,
compared to the requirements for various science cases at the GC.
The black points correspond to the reference frame precision during the era of Gaia-DR2,
the red points during the era of Gaia-DR3 (with the current value
represented by the red star), and the blue point is the predicted value that will be
achieved once Gaia-DR4 is released (expected sometime after 2026).}
\label{fig:ref_frame_science}
\end{figure}

The precision of the HST-Gaia reference frame enables its application to a wide range of science
cases at the GC.
The uncertainty in the proper motion bias, which represents the best possible
proper motion precision that can be obtained using the HST-Gaia reference frame,
has improved over time with the increasing number of HST epochs in the dataset
and with successive Gaia catalog releases (Figure \ref{fig:ref_frame_science}).
Today, the proper motion uncertainty achieved by the HST-Gaia reference frame
is within the requirements for science cases such as studying the dynamical structure of the NSC
and for probing the extended mass distribution around SgrA*.

The proximity of the NSC provides a unique opportunity to
study the formation and evolution of stars within a galactic nucleus in high detail.
Dynamical studies the NSC allows us to constrain its fundamental
properties such as its mass, shape, and rotation \citep[e.g.][]{Schodel:2009hh, Feldmeier:2014bk, Chatzopoulos:2015lq},
constrain possible formation mechanisms \cite[e.g.][]{Antonini:2012vy, Do:2020jr, Arca-Sedda:2020fc, Neumayer:2020nk},
and probe the relationship between
the NSC and other stellar structures at the GC \citep[e.g.][]{Nogueras-Lara:2022vo, Nogueras-Lara:2023wd}.
To probe the dynamical structure of the NSC,
it is necessary to measure stellar proper motions
at a precision that is significantly lower than the velocity dispersion of the cluster
\citep[$\sim$2.1 mas yr$^{-1}$ at the half-light radius;][]{Chatzopoulos:2015lq}.
To measure such a dispersion at 3$\sigma$ significance would thus require a
proper motion precision of $\sim$0.7 mas yr$^{-1}$, which is easily achieved by the HST-Gaia reference frame.

The population of stars and compact remnants at the GC results in
a distribution of extended mass around SgrA* that
yields insight into the many dynamical processes near a SMBH.
Processes such as dynamical relaxation, mass segregation, stellar collisions, SNe kicks, and tidal disruptions
modify the predicted extended mass distribution \citep[e.g.][]{Bahcall:1977wq, Alexander:2009eu, Alexander:2017ef, Baumgardt:2018pw, Linial:2022sx, Rose:2024vi, Zhang:2024cg, Jurado:2024gt}.
One approach to explore the extended mass is to measure the additional acceleration it
induces on the orbits of stars near the SMBH.
This requires a reference frame with a proper motion precision of approximately:

\begin{equation}
\sigma_{pm} \lesssim \frac{G M_{enc}}{r_{apo}} \frac{1}{t_{baseline}}
\end{equation}

where $M_{enc}$ is the total enclosed mass within the apocenter of the stellar orbit $r_{apo}$,
$t_{baseline}$ is the time baseline of the reference frame observations, and $G$ is the gravitational constant.
The current HST-Gaia reference frame delivers a minimum proper motion precision that
would allow for the detection of an extended mass of $\sim$3000 M$_{\odot}$ within the orbit of the star S0-2 ($r_{apo}$ = 0.01 pc),
and the future improvement from Gaia-DR4 is expected to extend that limit to $\sim$1000 M$_{\odot}$
(Figure \ref{fig:ref_frame_science}).
Thus, the HST-Gaia reference frame can be used to probe the range $M_{enc}$ values within S0-2 that are consistent
with current observational constraints \citep[1200 M$_{\odot}$ -- 5500 $M_{\odot}$ at the 1$\sigma$ confidence level;][]{Do:2019gr, GRAVITY-Collaboration:2022kk, The-GRAVITY-Collaboration:2024xt}.
Notably, HST-Gaia is the first ICRS-based reference frame to deliver the precision
required for this kind of analysis.

\section{Conclusions}
\label{sec:conclusions}
We combine multi-epoch HST observations of the GC with the Gaia-DR3
catalog to construct the first high-precision ICRS-based proper motion catalog of NIR astrometric reference sources
in the central parsec of the Galaxy.
New astrometric measurements are extracted from 14 epochs of HST WFC3-IR astrometry (2010 -- 2023)
and transformed into the Gaia-CRF3 reference frame using stars in common with Gaia-DR3.
Proper motions are measured using a new method that simultaneously models systematic errors
in the astrometry via Gaussian Processes.
This approach significantly improves the HST proper motion measurements in light of
a previously unknown systematic error discovered in the HST WFC3-IR astrometry.
We present a catalog of 2876 sources derived from this dataset:

\begin{itemize}
\item 40 Gaia-DR3 stars, the primary reference stars used to transform the HST astrometry into the Gaia-CRF3 coordinate system;
\item 2823 secondary reference stars within R $\lesssim$ 25'' (1 pc) from SgrA*, which overlap the region covered by long-term ground-based AO observations monitoring stellar orbits around the SMBH; and
\item 13 SiO masers, which are used to evaluate the consistency between HST-Gaia and radio proper motion measurements near the GC.
\end{itemize}

This catalog defines a novel HST-Gaia astrometric reference frame
that can applied to dynamical studies of stars near the central SMBH.

The HST-Gaia reference frame represents and extension of the ICRS coordinate system (via Gaia-CRF3) into
a region of the GC that is inaccessible through Gaia alone.
The HST-Gaia astrometric reference stars have measurement
uncertainties as low as $\sim$0.03 mas yr$^{-1}$ in proper motion and $\sim$0.11 mas in
reference epoch position, achieving a factor of $\sim$20x improvement over previous
ICRS proper motions published in the region \citep[][]{Griggio:2024dt}.
The uncertainty in the systematic bias between the HST-Gaia and Gaia-CRF3 reference
frames, which represents the best possible precision for HST-Gaia measurements,
is $\sigma_{pm}$ = 0.025 mas yr$^{-1}$ in proper motion and
$\sigma_{pos}$ = 0.044 mas in position.
This represents a factor of $\gtrsim$1.2x improvement in proper motion
and a factor of $\gtrsim$4x improvement in position relative to previous maser-based GC
astrometric reference frames in the NIR \citep{Plewa:2015ud, Sakai:2019fm}.
A key factor driving this improvement is the increased number of primary
reference stars available HST-Gaia (40 Gaia-DR3 sources)
compared to previous GC NIR reference frames (7 -- 8 stellar SiO masers).

In addition, HST-Gaia is the first GC NIR reference frame that is independent of
radio measurements of stellar SiO masers near SgrA*.
This enables tests for systematics between the HST-Gaia and radio measurements.
We compare the positions and proper motions for 13 stellar masers
found in both the HST-Gaia catalog and the most recent radio maser reference
frame catalog from \citetalias{Darling:2023ao}.
After transforming the HST-Gaia measurements into a SgrA*-at-Rest coordinate system
(and accounting for the additional uncertainties incurred),
we find that the average difference between the HST-Gaia and radio measurements
are (-0.076 $\pm$ 0.030, -0.042 $\pm$ 0.028) mas yr$^{-1}$ in the ($\mu_{\alpha*}^s$, $\mu_{\delta}^s$)
proper motions and (-0.536 $\pm$ 0.231, -0.517 $\pm$ 0.486) mas in the ($\delta_{\alpha*}^s$, $\delta_{\delta}^s$) positions.
This indicates that the HST-Gaia and radio reference frames are
consistent to within 0.041 mas yr-1 in proper motion and 0.54 mas in position at 99.7\% confidence.
There is a possible tension in the $\alpha^*$ direction at $\sim$2.5$\sigma$ significance,
but further observations are required to determine if it is indeed real and what its source(s) might be.

Moving forward, the HST-Gaia reference frame can be applied to
a range of dynamical science cases at the
GC, such as measuring the internal kinematics of the NSC and constraining
the enclosed mass within the orbits of stars near SgrA*.
Future improvements will be achieved with the release of
the Gaia-DR4 catalog (scheduled for sometime after mid-2026),
which we predict will further reduce the overall uncertainty
in the HST-Gaia reference frame by an additional factor of $\sim$2x.
The complete proper motion catalog of stars derived from the HST-Gaia
observations will be published in a future paper (Hosek et al., in prep).
This rich dataset will establish a lasting legacy on the dynamical studies of stars near the GC.

\acknowledgements
The authors thank the anonymous referee for their feedback which improved the paper. M.W.H. is supported by the Brinson Prize Fellowship, and the authors acknowledge support from the Gordon E. \& Betty I. Moore Foundation award \#11458. This work is based on observations made with the NASA/ESA Hubble Space Telescope, obtained at the Space Telescope Science Institute, which is operated by the Association of Universities for Research in Astronomy, Inc., under NASA contract NAS 5-26555. The observations are associated with programs GO-11671, GO-12318, GO-12667, GO-13049, GO-15199, GO-15498, GO-16004, GO-15894, GO-16681, and GO-16990. This research has made extensive use of the NASA Astrophysical Data System.

\facilities{HST (WFC3-IR), Gaia}

\software{AstroPy \citep{Astropy-Collaboration:2013kx}, texttt{img2xym\_wfc3ir} \citep[a precursor to \texttt{hst1pass} described in][]{Anderson:2022vs},  \texttt{KS2} \citep{Anderson:2008qy}, Matplotlib \citep{Hunter:2007}, numPy \citep{oliphant2006guide}, SciPy \citep{Virtanen:2020fp}}

\clearpage

\appendix

\section{A Systematic Error in HST WFC3-IR Astrometry With Position Angle}
\label{app:ast_errs}
\begin{figure*}
\begin{center}
\includegraphics[scale=0.3]{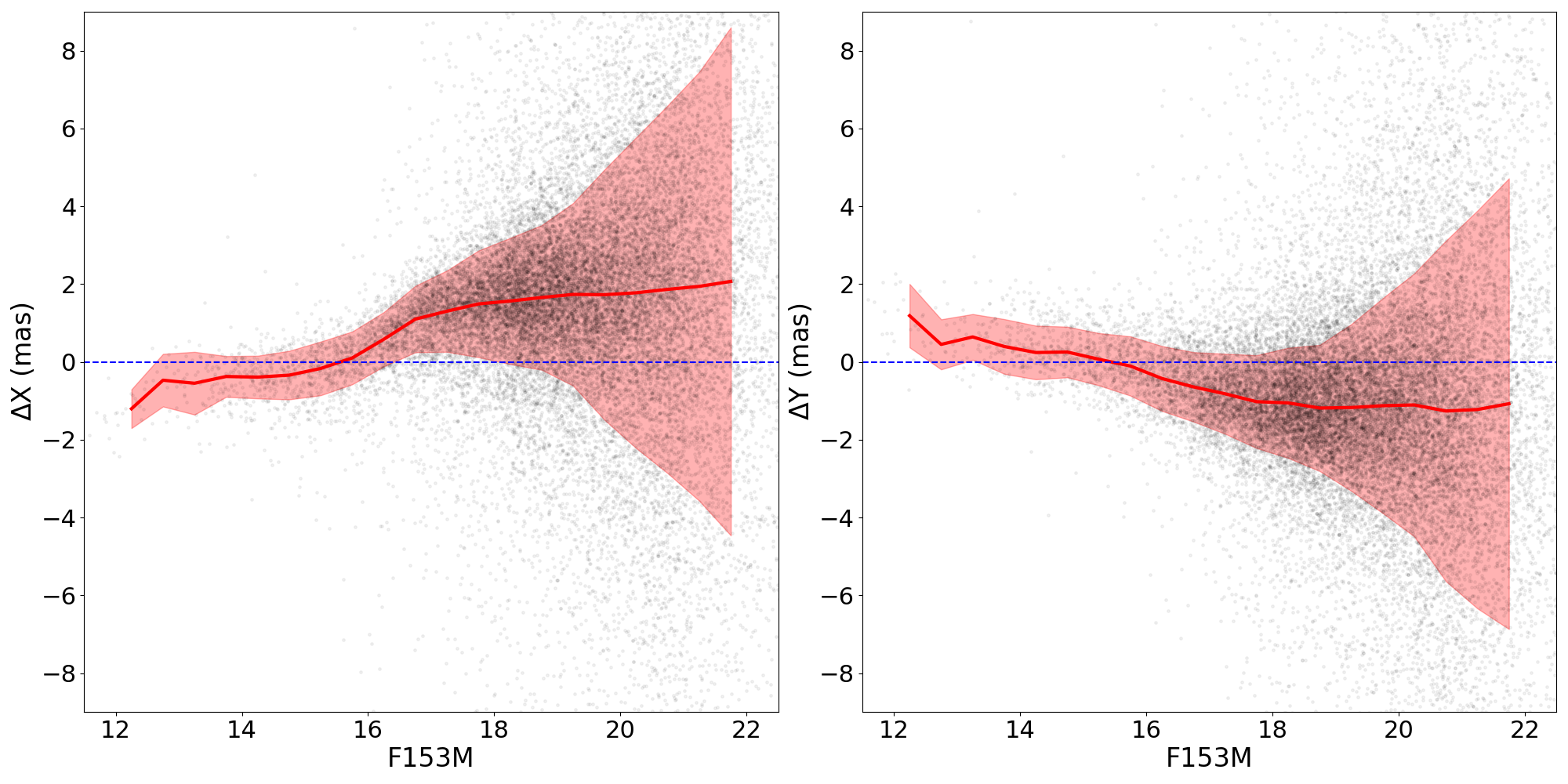}
\caption{Systematic difference in HST astrometry measured at PA = -45$^{\circ}$ and PA = 134$^{\circ}$ epochs as a function of magnitude,
in detector X and Y coordinates (left and right panels, respectively). The red line and shaded region shows the average and standard deviation of the position difference,
which reaches $\sim$2 mas for fainter stars.
}
\label{fig:pos_resids}
\end{center}
\end{figure*}

Due to restrictions on the HST orbital visibility, the GC observations were taken at
two PAs separated by 180 degrees depending on the time of year.
A systematic offset in the HST astrometry was found between observations taken
at these different PAs.
The size of the offset increases as a function of magnitude such that
the positions of fainter stars shift by $\sim$2 mas in both the X and Y
directions on the detector (Figure \ref{fig:pos_resids}).
When rotated onto the plane of the sky, this shift manifests almost entirely the DEC direction.

\begin{figure}
\begin{center}
\includegraphics[scale=0.35]{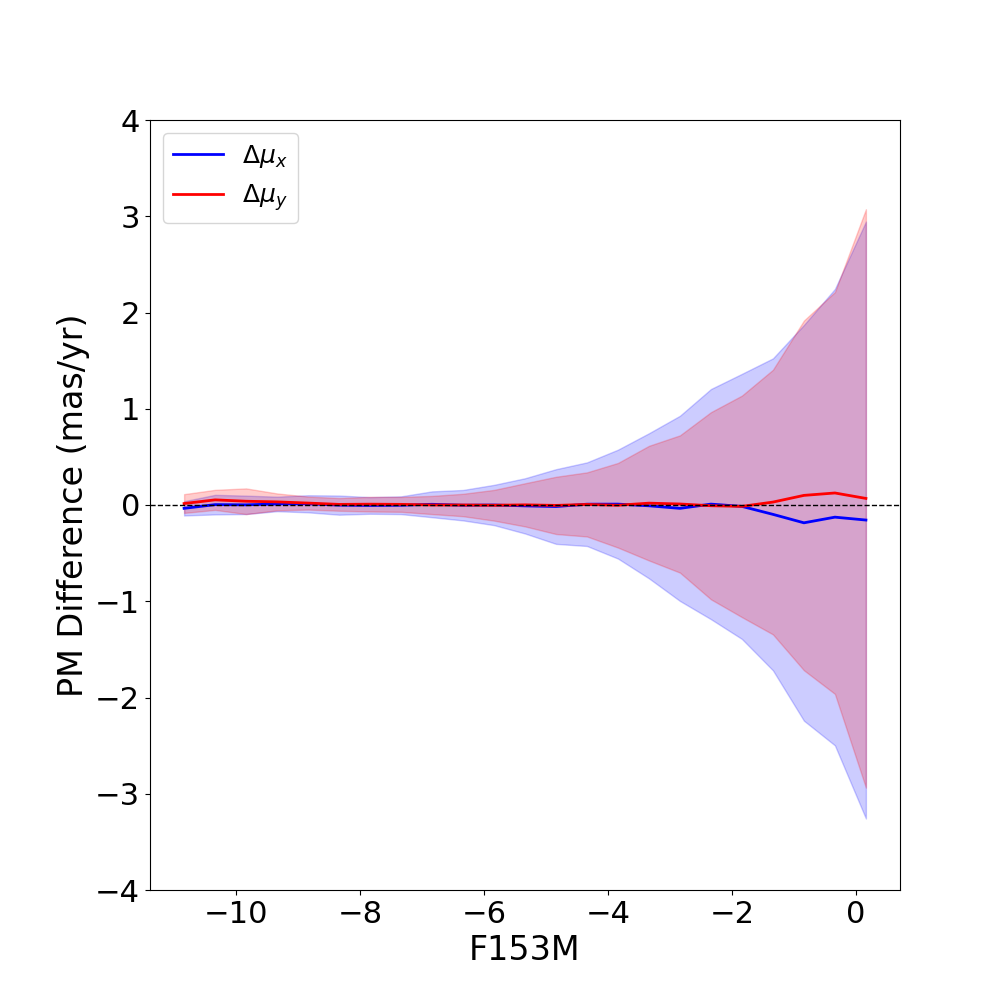}
\caption{The difference between proper motions measured from PA = -45$^{\circ}$ and PA = 134$^{\circ}$ epochs independently as a function of mag.
The red and blue lines and corresponding shaded regions correspond to the average and standard deviation of the differences in the detector X and Y directions,
respectively. There are no significant differences in the proper motions, suggesting that the systematic position offset is constant in amplitude and direction (relative to the detector) across the dataset.}
\label{fig:pm_PA_comp}
\end{center}
\end{figure}

The offset appears to remain constant over the time baseline of the observations,
as linear proper motions derived from only the PA = -45$^{\circ}$ epochs (8 epochs between 2010.6261 -- 2023.6178)
and only the PA = 134$^{\circ}$ epochs (6 epochs between 2014.099 -- 2023.1535)
are fully consistent within uncertainties (Figure \ref{fig:pm_PA_comp}).

The cause of this systematic offset is not yet clear.
If it were caused by uncorrected optical distortion,
then we would expect the position shift to be a function
of location on the detector rather than a function of magnitude,
since bright and faint stars would presumably be equally impacted by the distortion.
Similarly, if there were a bias in the astrometric
reference frame between the different PAs (for example, if induced by parallax in the reference stars),
then we would also expect it to impact both bright and faint stars equally.

We hypothesize that
this systematic may be explained by a drift in the HST pointing
over the length of the individual exposures (349 seconds).
All F153M astrometric observations were obtained using the
STEP50 up-the-ramp sampling sequence.
In this setup, the positions of bright stars come from an average of the
first few detector reads before non-linearity/saturation is reached,
while the positions of fainter stars come from
averaging all the reads across the exposure.
Thus, a drift in the telescope pointing during an exposure would result in an apparent position
shift between the bright and faint stars \citep[][]{Anderson:2022vs}.
If the telescope drift were consistent in both direction and amplitude
(relative to detector coordinates) across the dataset,
then the position offset would only become apparent when the observation PA is changed,
since the direction of the position shift on the sky would change.
However, further investigation into this systematic is required
to definitively establish its origin.

\section{Modeling polynomial motion using Gaussian Processes}
\label{app:gpoly}

Gaussian processes are a powerful and generic statistical tool that are used to model, parameterically or non-parameterically, a wide-range of physical processes \citep{Rasmussen:2006zk}.  For our purposes, we define a Gaussian Process as a process whose predicted observable is described by a linear model ($\mat{f}(\mat{t}) = \mat{A}(\mat{t}) \mat{w}$), and whose measured observables ($\mat{x}$) and parameters ($\mat{w}$) (prior probability densities) are normally distributed:
\begin{eqnarray}
\label{eq:1}
\mat{x}\mid\mat{w} & \sim & \mathcal{N}(\mat{A}(\mat{t})\cdot\mat{w}, \mat{\Sigma}) \\
\mat{w} & \sim & \mathcal{N}(\mat{0}, \mat{\Sigma}_p).
\end{eqnarray}
Note that the matrix $\mat{A}(\mat{t})$ can be any function of any set of conditional parameters, $\mat{t}$.  Here we impose that $\mat{w}$ is distributed about $\mat{0}$ as to minimize the effect of unconstrained parameters on the resultant posterior distributions.  Under these assumptions, the marginal likelihood (marginalized over $\mat{w}$) is
\begin{equation}
\label{eq:1b}
\mat{x} \mid \mat{t} \sim \mathcal{N}\left(\mat{0},\, \mat{\Sigma} + \mat{K}(\mat{t}, \mat{t})\right)
\end{equation}
and any predicted observable corresponding to $\mat{t}_{\mathrm{p}}$ ($\mat{f}(\mat{t}_{\mathrm{p}}) = \mat{A}(\mat{t}_{\mathrm{p}})\mat{w}$) is normally distributed with a mean and variance of
\begin{subequations}
\label{eq:margcondp}
\begin{eqnarray}
\ex[\mat{w}\mid \mat {x}, \mat{t}]{\mat{f}(\mat{t}_{\mathrm{p}})} & = &
\mat{K}(\mat{t}_{\mathrm{p}}, \mat{t})\left(\mat{\Sigma} + \mat{K}(\mat{t}, \mat{t})\right)^{\minus 1}\mat{x} \\
\var[\mat{w}\mid \mat {x}, \mat{t}]{\mat{f}(\mat{t}_{\mathrm{p}})} & = &
\mat{K}(\mat{t}_{\mathrm{p}}, \mat{t}_{\mathrm{p}}) - \mat{K}(\mat{t}_{\mathrm{p}}, \mat{t})\left(\mat{\Sigma} + \mat{K}(\mat{t}, \mat{t})\right)^{\minus 1} \mat{K}(\mat{t}, \mat{t}_{\mathrm{p}})
\end{eqnarray}
\end{subequations}
where $\ex[\mat{x}\mid \ldots]{\mat{f}(\mat{x})} \equiv \int \mat{f}(\mat{x}) \mathcal{P}(\mat{x}\mid \ldots) \mathrm{d}\mat{x}$ is the expectation,  $\var[\mat{x}\mid \ldots]{\mat{f}(\mat{x})} \equiv \int \left(\mat{f}(\mat{x})\right)^2 \mathcal{P}(\mat{x}\mid \ldots) \mathrm{d}\mat{x}$ is the variance, and $\mat{K}(\mat{t}, \mat{t}') = \mat{A}(\mat{t})\mat{\Sigma}_p\mat{A}^{\T}(\mat{t}')$ is defined as the kernel.  Or equivalently, the kernel can be expressed as a mean and covariance of $\mat{f}(\mat{t})$:
\begin{eqnarray}
m(\mat{t}) & = & \ex[\mat{w}]{\mat{f}(\mat{t};\, \mat{w})}\label{eq:meandef}\\
K(\mat{t}, \mat{t}') & = & \ex[\mat{w}]{\left(\mat{f}(\mat{t};\, \mat{w})-m(\mat{t})\right)\left(\mat{f}(\mat{t}';\, \mat{w})-m(\mat{t}')\right)} \label{eq:kerneldef}
\end{eqnarray}

Note that here, by convention, $m(\mat{t}) = 0$.

\subsection{polynomial kernels}
\label{app:kernels:poly}

Since polynomial motion is linear with respect to the polynomial moments, a $N$-th order polynomial motion can be described by a Gaussian Process where
\begin{eqnarray}
\mat{f}(\mat{t}) &=& x_0 + v(\mat{t} - t_0) + \frac{a}{2}(\mat{t}-t_0)^2 + \frac{j}{6}(\mat{t}-t_0)^3 + ... \nonumber \\
&=& \mat{A}  \mat{w} \\
\mat{w} & \equiv & [x_0, v, a, j, ...]^\T \\
\mat{t} & = & [t_1, t_2, \ldots]^\T
\end{eqnarray}
and
\begin{equation}
\mat{A} = \left[\begin{array}{cccccc}
1 & (t_1 - t_0) & \frac{1}{2}(t_1 - t_0)^2 & \frac{1}{6}(t_1 - t_0)^3  & \cdots & \frac{1}{N!}(t_1-t_0)^N \\
1 & (t_2 - t_0) & \frac{1}{2}(t_2 - t_0)^2 & \frac{1}{6}(t_2 - t_0)^3 & \cdots & \frac{1}{N!}(t_2-t_0)^N \\
\vdots & \vdots & \vdots & \vdots &  & \vdots
\end{array}\right].
\end{equation}
With little loss in generality, we assume $\mat{\Sigma}_p$ is diagonal, $\Sigma_{p, m n} = \sigma_n^2\delta_{m n}$.  With this, the Kernel element for a polynomial becomes:
\begin{equation}
K^{\mathrm{(poly)}}(t, t') = \sum_{n=0}^N \frac{\sigma_n^2}{(n!)^2}\left[(t-t_0)(t'-t_0)\right]^n.
\end{equation}
Depending on the parameterization of $\mat{\sigma}_p$, the polynomial kernel reduces to:
\begin{equation}
\label{eq:B12}
K^{\mathrm{(poly)}}(t, t') = \left\{\begin{array}{ll}
\sigma_0^2\exp\left[\left(\frac{t-t_0}{\tau}\right)\left(\frac{t'-t_0}{\tau}\right)\right]; & \begin{array}{l}\sigma_n^2 = n!\left(\frac{\sigma_0}{\tau^{n}}\right)^2 \\ N \rightarrow \infty \end{array} \\
\sigma_0^2\left[1+\left(\frac{t-t_0}{\tau}\right)\left(\frac{t'-t_0}{\tau}\right)\right]^N; & \sigma_n^2 = \frac{n!\, N!}{(N-n)!}\left(\frac{\sigma_0}{\tau^{n}}\right)^2
\end{array}\right.
\end{equation}

Note that the top and bottom parameterizations in Equation \ref{eq:B12} correspond to an infinite and finite polynomial, respectively.

\subsection {polynomial moment predictions}
\label{app:poly_higher_order_pred}
In our case, we are interested in constraining the moments of the polynomial.
Any order moment of the polynomial can be directly inferred from Equation \ref{eq:margcondp}, with the understanding that when calculating higher order moments, derivatives are applied only to the polynomial component, $f^{(n)}(t) \equiv \mathrm{d}^n f(t) / \mathrm{d}t^n = \mathrm{d}^n f^{\mathrm{(poly)}}(t) / \mathrm{d} t^n$.  Since any order polynomial moment is a linear model, its prediction is also normally distributed with the mean and covariance of
\begin{eqnarray}
\ex[\mat{w} \mid \mat{x}]{f^{(n)}(t')} & \equiv & \ex[\mat{w}\mid \mat{x}]{\left.\frac{\mathrm{d}^n}{\mathrm{d}\tilde{t}^n} f\left(\tilde{t}\right)\right|_{\tilde{t}=t'}} \\
& = & \left.\frac{\mathrm{d}^n}{\mathrm{d}\tilde{t}^n}\ex[\mat{w} \mid \mat{x}]{f\left(\tilde{t}\right)}\right|_{\tilde{t}=t'} \\
& = & \left.\frac{\partial^n}{\partial \tilde{t}^n}K^{\mathrm{(poly)}}\left(\tilde{t}, \mat{t}\right)\right|_{\tilde{t} = t'} \left(\mat{\Sigma} + \mat{K}(\mat{t}, \mat{t})\right)^{\minus 1}\mat{x}
\end{eqnarray}
and
\begin{eqnarray}
\cov[\mat{w} \mid \mat{x}]{f^{(m)}(t')}{f^{(n)}(t')}
& \equiv & \cov[\mat{w} \mid \mat{x}]{\left.\frac{\mathrm{d}^m}{\mathrm{d} \tilde{t}^m}f\left(\tilde{t}\right)\right|_{\tilde{t}=t'}}{\left.\frac{\mathrm{d}^n}{\mathrm{d} \tilde{t}'^n}f\left(\tilde{t}'\right)\right|_{\tilde{t}'=t'}} \\
& = & \left.\frac{\mathrm{d}^m}{\mathrm{d} \tilde{t}^m} \frac{\mathrm{d}^n} {\mathrm{d} \tilde{t}'^n}\cov[\mat{w} \mid \mat{x}]{f\left(\tilde{t}\right)}{ f\left(\tilde{t}'\right)}\right|_{\tilde{t}=t', \tilde{t}'=t'} \\
& = & \left.\frac{\partial^m}{\partial \tilde{t}^m} \frac{\partial^n}{\partial \tilde{t}'^n} K^{\mathrm{(poly)}}\left(\tilde{t}, \tilde{t}'\right)\right|_{\tilde{t}=t', \tilde{t}'=t'} \nonumber \\
& &- \left.\frac{\partial^m}{\partial \tilde{t}^m}K^{\mathrm{(poly)}}\left(\tilde{t}, \mat{t}\right)\right|_{\tilde{t} = t'} \left(\mat{\Sigma} + \mat{K}(\mat{t}, \mat{t})\right)^{\minus 1}\left.\frac{\partial^n}{\partial \tilde{t}'^n}K^{\mathrm{(poly)}}\left(\mat{t}, \tilde{t}'\right)\right|_{\tilde{t}' = t'}.
\end{eqnarray}
Here, $\cov[\mat{x} \mid \ldots]{f(\mat{x})}{ g(\mat{x})} \equiv \int \left(f(\mat{x}) - \ex[\mat{x} \mid \ldots]{f(\mat{x})}\right) \left(g(\mat{x}) - \ex[\mat{x} \mid \ldots]{g(\mat{x})}\right) \mathcal{P}(\mat{x} \mid \ldots) \mathrm{d}\mat{x}$ is the covariance.

\subsection{time correlated kernels}
\label{app:kernels:timecorr}
Systematics such as those in Appendix \ref{app:ast_errs} manifest as positional shifts in time.  Here, we model this effect as a shift within some window of width $\tau$ centered at time $t_0$.  To account for the varying position position shifts these systematics cause at different times, several window functions are used, each weighted independently:
\begin{equation}
f(t; t_0) = \sum_i w_i W(t - t_{0, i})
\end{equation}
For generality, $\{t_{0, i}\}$, and its associated weights, $\{w_i\}$, can be described as a grid with spacing of $\Delta t_0$, where the normal priors of $\{w_i\}$ centered at $zero$ minimizes the effect of any unconstrained $t_{0, i}$.  If an infinitely large grid of infinitesimal spacing ($\Delta t_0 \rightarrow 0$) is assumed, the Kernel for this process can be reduced to
\begin{eqnarray}
K(t, t') & = & \sum_i \sigma^2_i W(t - t_{0, i}) W(t' - t_{0, i}) \\
& = & \sigma^2_0 \sum_i \frac{\Delta t_0}{Z}W(t - t_{0, i}) W(t' - t_{0, i}) \\
& \approx & \sigma^2_0 \int \frac{\mathrm{d} t_0}{Z} W(t - t_0) W(t' - t_0)
\end{eqnarray}
where the covariance of the prior on $\mat{w}$ is assumed to be diagonal with values of $\sigma_i^2 \equiv \Sigma_{p, ii} = \sigma^2_0 \Delta t_0 / Z$ and $Z = \int W(t)^2 \mathrm{d} t$.  Here, we use two window functions, a step function and a squared exponential window function:
\begin{eqnarray}
W^{\mathrm{(step)}}(t) &=& \left\{
\begin{array}{cl}
1 & \mathrm{\ for\ } \left|t\right| \leq \frac{\tau}{2} \\
0 & \mathrm{\ for\ } \left|t\right| > \frac{\tau}{2}
\end{array}
\right. \\
W^{\mathrm{(sqexp)}}(t) &=& \exp\left[-\frac{t^2}{\tau^2}\right]
\end{eqnarray}
These lead to the kernels:
\begin{equation}
K^{\mathrm{(step)}}(t, t') = \sigma^2_0 \times \left\{
\begin{array}{cl}
1 - \left|\frac{t - t'}{\tau}\right| & \mathrm{for\ } \left|t - t'\right| \leq \tau \\
0 & \mathrm{for\ } \left|t - t'\right| > \tau
\end{array} \right.
\end{equation}
and
\begin{equation}
K^{\mathrm{(sqexp)}}(t, t') = \sigma_0^2 \exp\left[-\frac{(t-t')^2}{2 \tau^2}\right].
\end{equation}

\subsection{spatial correlated kernels}
\label{app:kernels:spatial}
Systematics due to confusion with unresolved point sources can manifest as spatially correlated observations.  Again, we model this effect as a shift weighted by some window of width $\ell$ with the exception that this shift is toward some position $\mat{r}_0 = \{x_0, y_0\}$.  Again by assuming a grid of $\mat{r}_c$, each with its own weight, leads to a perturbation of
\begin{equation}
f(\mat{r}; \mat{r}_0) = \sum_i w_i (\mat{r} - \mat{r}_{0, i}) W(\mat{r} - \mat{r}_{0, i}).
\end{equation}
The kernel as a function of the real spatial coordinates, $\mat{r}$ can be shown to be
\begin{eqnarray}
K(\mat{r}, \mat{r}') & = & \sum_i \sigma^2_i (\mat{r} - \mat{r}_{0, i})(\mat{r}' - \mat{r}_{0, i}) W(\mat{r} - \mat{r}_{0, i}) W(t' - t_{0, i}) \\
& = & \sigma^2_0 \sum_i \frac{\Delta \mat{r}_0}{Z} (\mat{r} - \mat{r}_{0, i})(\mat{r}' - \mat{r}_{0, i}) W(\mat{r} - \mat{r}_{0, i}) W(\mat{r}' - \mat{r}_{0, i}) \\
& \approx & \sigma^2_0 \int \frac{\mathrm{d} \mat{r}_0}{Z} (\mat{r} - \mat{r}_0)(\mat{r}' - \mat{r}_0) W(\mat{r} - \mat{r}_0) W(\mat{r}' - \mat{r}_0)
\end{eqnarray}
where the covariance of the prior on $\mat{w}$ is assumed to be diagonal with values of $\sigma_i^2 \equiv \Sigma_{p, ii} = \sigma^2_0 \Delta \mat{r}_0 / Z$ and $Z = \int \mat{r}^2 W(\mat{r})^2 \mathrm{d} \mat{r}$.  For an squared exponential window function of
\begin{equation}
W(\mat{r}) = \exp\left[-\frac{\left|\mat{r}\right|^2}{\ell^2}\right],
\end{equation}
the kernel becomes:
\begin{equation}
K(\mat{r}, \mat{r}') = \sigma_0^2 \left[\mat{1} - \frac{(\mat{r} - \mat{r}')(\mat{r} - \mat{r}')}{\ell^2}\right] \exp\left[-\frac{\left|\mat{r} - \mat{r}'\right|^2}{2\ell^2}\right]
\end{equation}
Unfortunately, $r$, and thus the dispersion, is dependant on the polynomial motion being modeled, making marginalization of the polynomial parameters non-trivial.  Here we make a normal approximation of the marginal probability density by setting resultant variance using Equation \ref{eq:kerneldef}:
\begin{eqnarray}
K^{\mathrm{(conf)}}(t, t') & = & \ex[w_t, w_r]{\mat{r}(t)\mat{r}(t')^{\T}} \\
& = & \ex[w_t]{K(\mat{r}(t), \mat{r}(t'))}.
\end{eqnarray}
Note that here the mean evaluates to zero.  This results in a dispersion of:
\begin{equation}
K^{\mathrm{(conf)}}(t, t') =  \sigma_0^2 \frac{\left[\mat{1} + \frac{K^{\mathrm{(tot)}}(t, t) + K^{\mathrm{(tot)}}(t', t') - 2\mat{K}^{\mathrm{(tot)}}(t, t')}{\ell^2}\right]^{-1}}{\sqrt{\left|\mat{1} + \frac{K^{\mathrm{(tot)}}(t, t) + K^{\mathrm{(tot)}}(t', t') - 2 K^{\mathrm{(tot)}}(t, t')}{\ell^2}\right|}}
\end{equation}
where $\mat{K}^{\mathrm{(tot)}}$ is the full covariance matrix in $x$ and $y$.
\begin{equation}
K^{\mathrm{(tot)}}(t, t') = \left[\begin{array}{cc} K_x(t, t') & 0 \\ 0 & K_y(t, t') \end{array}\right].
\end{equation}
and $K_x(t, t')$ and $K_y(t, t')$ are the polynomial and time-dependent covariances in the $x$ and $y$ directions.

\subsection{hyper-parameters fitting and penalty term}

For the above kernels, we fit the hyper-parameters, $\ell$, $\tau$, and $\sigma_0^2$, by minimizing the marginal likelihood (Equation \ref{eq:1b}).  Since the Gaussian Processes model physical effects as correlations, any unaccounted for temporal auto-correlations in the data can bias the results.  This has the largest affect on $\tau$ and $t_0$, usually causing these quantities to sometimes diverge.  For this reason, we set $t_0$ to the weighted average and impose a prior on $\tau$ of
\begin{equation}
\mathcal{P}(\tau) \propto \frac{1}{1 + \tau}.
\end{equation}

\subsection{leave-one-out cross-validation (LOO-CV) and the expected-log-probablity-density (ELPD)}\label{app:loocv}

Leave-one-out cross-validation (LOO-CV) compares predicted results to a subset of data, $\mat{x}_k$, when the model is fitted to a data set that excludes this subset, $\mat{x}_{\minus k} = \{\mat{x}_0, \ldots, \mat{x}_{k-1}, \mat{x}_{k+1}, \ldots\}$ ($\mathcal{P}(\mat{x}_k \mid \mat{x}_{\minus k})$). If the hyper-parameters are not expected to change a significant amount when one data point is removed, then the variance in the marginalized parameters, $\mat{w}$, will dominate. Because in a Gaussian process model the predicted values are also normally distributed, one can calculate the predicted value of a single data point ($x_{\mathrm{loo}_k}$) and variance of that point ($\mat{\sigma}^2_{\mathrm{loo}_k}$):
\begin{eqnarray}
x_{\mathrm{loo}_k} & = & \mat{K}(t_k, \mat{t}_{\minus k}) \left(\mat{\Sigma}_{-k} + \mat{K}(\mat{t}_{\minus k}, \mat{t}_{\minus k})\right)^{-1} \mat{x}_{\minus k}\\
\mat{\sigma}^2_{\mathrm{loo}_k} & = & \mat{\Sigma}_k + \mat{K}(t_k, t_k) - \mat{K}(t_k, \mat{t}_{\minus k}) \left(\mat{\Sigma}_{\minus k} + \mat{K}(\mat{t}_{\minus k}, \mat{t}_{\minus k})\right)^{-1} \mat{K}(\mat{t}_{\minus k}, t_k)
\end{eqnarray}
(see Equation \ref{eq:margcondp}).  Here, $\mat{\Sigma}$ is assume to diagonal with elements $\{\mat{\Sigma}_k\}$. The leave-one-out analysis can also be used to calculate the Expect-Log-Probability-Density (Equation \ref{eq:elpd}):
\begin{eqnarray}
\mathrm{ELPD} & \equiv & \int \mathcal{P}(\mat{x}^*) \log \mathcal{P}(\mat{x}^* \mid \mat{x})\mat{d} \mat{x}^* \\
& \approx & \sum_k \log \mathcal{P}(\mat{x}_k \mid \mathbf{x}_{\minus k}) \\
& = & \sum_k \log \mathcal{N}(x_k \mid x_{\mathrm{loo}_k}, \sigma^2_{\mathrm{loo}_k})
\end{eqnarray}

\section{Transforming HST-Gaia Measurements into the SgrA*-at-Rest Coordinate System: Impact of Different ICRS Measurements of SgrA*}
\label{app:sgra_motion}
The ICRS position and proper motion of SgrA* is required in
order to transform the HST-Gaia measurements into a SgrA*-at-Rest frame ($\mathsection$\ref{sec:sgra_rest}).
Throughout this paper we adopt the ICRS measurements of SgrA* from \citet{Xu:2022sx},
but other recent measurements are available from
\citet{Reid:2020jo}\footnote{\citet{Reid:2020jo} measure the ICRS position of SgrA* to be
$\alpha$(J2000) = 17$^h$45$^m$40$^s$.0409 $\pm$ 0$^s$.011 and $\delta$(J2000) = $-$29$^{\circ}$00$'$28.$''$118 $\pm$ 0$^s$.01
at a reference epoch of 1996.215 \citep[see also][]{Reid:2004xh}
with a proper motion of $ \mu_{\alpha^*}$ = -3.156 $\pm$ 0.006 mas yr$^{-1}$,  $\mu_{\delta}$ = -5.585 $\pm$ 0.010 mas yr$^{-1}$.} and
\citet{Gordon:2023ck}\footnote{\citet{Gordon:2023ck} measure the ICRS position of SgrA* to be
$\alpha$(J2000) = 17$^h$45$^m$40$^s$.034047 $\pm$ 0$^s$.000018,
$\delta$(J2000) = $-$29$^{\circ}$00$'$28.$''$21601 $\pm$ 0$''$.00044 at a reference epoch of 2015.0 with a proper motion of
$ \mu_{\alpha^*}$ = -3.128 $\pm$ 0.042 mas yr$^{-1}$,  $\mu_{\delta}$ = -5.584 $\pm$ 0.075 mas yr$^{-1}$.}.
All three of these studies measure the astrometry of SgrA* relative to background extragalactic sources
via multi-epoch Very Long Baseline Array (VLBA) observations.
\citet{Reid:2020jo} measure the position of SgrA* relative to two extragalactic sources
within 1$^{\circ}$ of SgrA*,
\citet{Xu:2022sx} by using four nearby extragalactic sources, the same two sources as \citet{Reid:2020jo} plus
two additional sources at larger radii from SgrA* ($\sim$2$^{\circ}$),
and \citet{Gordon:2023ck} by performing a global fit of the position of SgrA* using
over 1000 extragalactic sources observed across the sky.

\begin{figure*}
\includegraphics[scale=0.35]{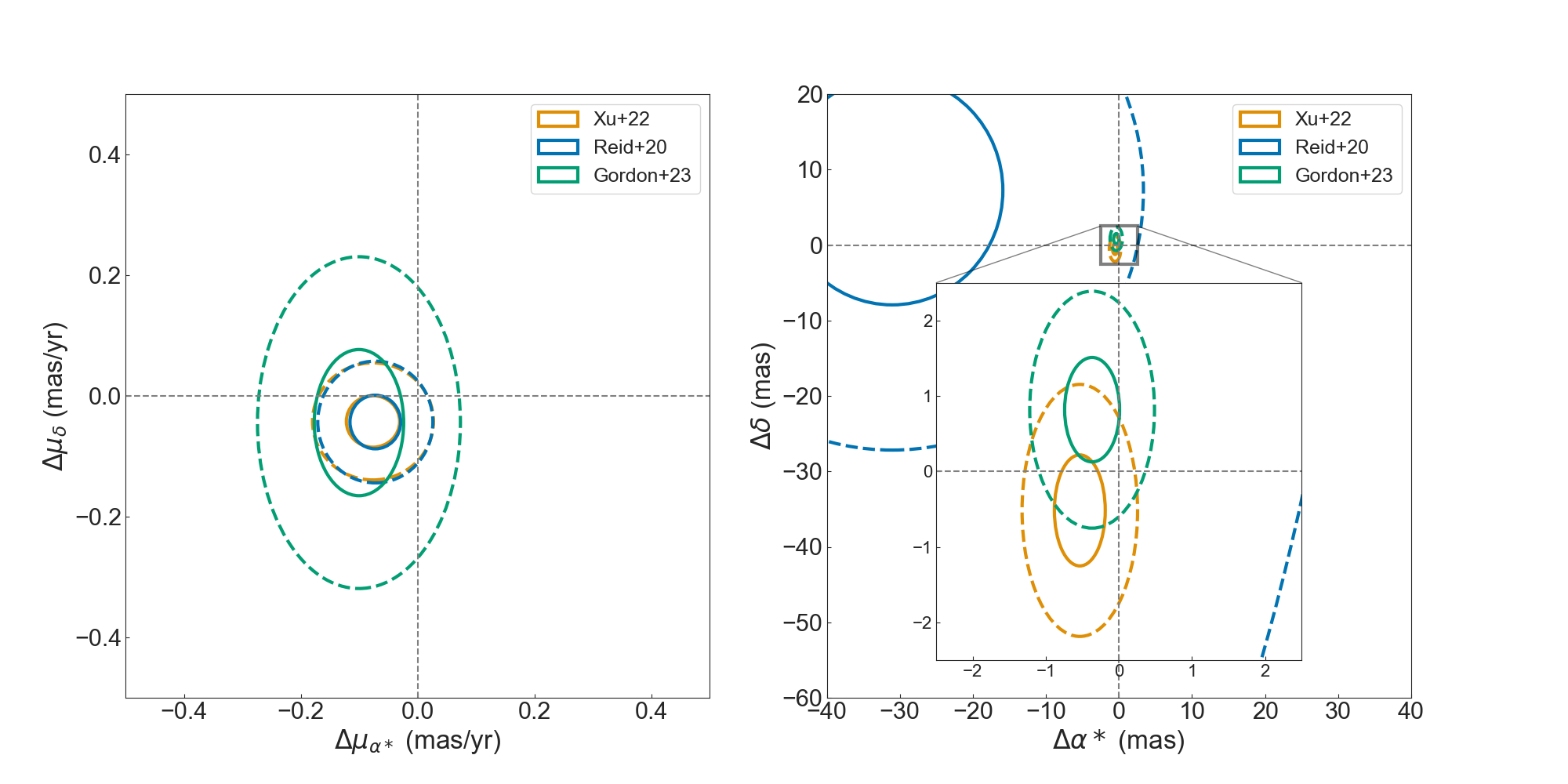}
\caption{The impact of choosing different ICRS positions and proper motions for SgrA* on the comparison between the HST-Gaia and radio
measurements of the stellar masers near SgrA* (e.g. $\mathsection$\ref{sec:hst_gaia_masers}).
The error-weighted mean difference in the maser proper motions (left) and positions (right)
are shown using the ICRS SgrA* values from \citet{Xu:2022sx} (orange), \citet{Reid:2020jo} (blue), and \citet{Gordon:2023ck} (green).
The solid and dashed lines correspond to the 68.3\% and 99.7\%
confidence level of the mean differences, respectively. The \citet{Xu:2022sx} measurement is adopted for this paper, which has
the smallest proper motion uncertainties and removes the $\sim$30 mas position offset that occurs when the \citet{Reid:2020jo} measurement is used.
}
\label{fig:sgra_comp}
\end{figure*}

Changing the ICRS position and proper motion of SgrA* can impact the
comparison between the HST-Gaia and radio measurements for the stellar
masers performed in $\mathsection$\ref{sec:hst_gaia_masers}.
Here we repeat the analysis from this section using the SgrA* ICRS
measurements from \citet{Reid:2020jo} and \citet{Gordon:2023ck} and compare
to the results obtained when the \citet{Xu:2022sx} measurement is used.
The error-weighted mean differences in the positions and proper motions of the
masers and their corresponding uncertainties are shown in Figure \ref{fig:sgra_comp}
and reported in Table \ref{tab:sgra_comp_summary}.
In all cases, the HST-Gaia SgrA*-at-Rest values are consistent with the
\citetalias{Darling:2023ao} radio measurements at the 99.7\% confidence level,
indicating that the HST-Gaia and radio reference frames are consistent in all cases.

The choice of SgrA* ICRS measurement has little impact on the proper motions,
as the proper motions across the studies are consistent within $\sim$0.024 mas yr$^{-1}$.
The uncertainty in the \citet{Gordon:2023ck} SgrA* proper motion is slightly higher than
that of \citet{Xu:2022sx} and \citet{Reid:2020jo}, which results in a slightly larger uncertainty in the
reference frame.
The positions show larger differences, most noticeably with
the \citet{Reid:2020jo} SgrA* position introducing a
$\sim$30 mas offset relative to the other studies, albeit with a large uncertainty.
This supports the conclusion that the \citet{Reid:2020jo}
position value is biased due to scatter broadening of radio sources within 1$^{\circ}$, where the extragalactic
reference sources used in that study are located.
Both \citet{Xu:2022sx} and \citet{Gordon:2023ck} establish the position of SgrA* using reference sources beyond the
large scatter broadening region and thus are more accurate.
For this paper, we adopt the SgrA* ICRS position and proper motion from \citet{Xu:2022sx}
as it has lower proper motion uncertainties than \citet{Gordon:2023ck} and
resolves the position offset from \citet{Reid:2020jo}.

\begin{deluxetable*}{l c c c c}
\tabletypesize{\footnotesize}
\label{tab:sgra_comp_summary}
\tablecaption{Consistency Between HST-Gaia and Radio Masers When Adopting Different ICRS Measurements of SgrA*}
\tablehead{
\colhead{Ref} & \colhead{$\Delta\delta_{\alpha*}^s$} & \colhead{$\Delta\delta_{\delta}^s$} & \colhead{$\Delta\mu_{\alpha*}^s$} &  \colhead{$\Delta\mu_{\delta}^s$} \\
&  (mas)  & (mas) & (mas yr$^{-1}$) & (mas yr$^{-1}$)
}
\startdata
\citet{Xu:2022sx}\tablenotemark{a} & -0.536 $\pm$ 0.230 & -0.517 $\pm$ 0.486 & -0.076 $\pm$ 0.030 & -0.042 $\pm$ 0.028 \\
\citet{Reid:2020jo} & -31 $\pm$ 10 & 7 $\pm$ 10 & -0.072 $\pm$ 0.029 & -0.043 $\pm$ 0.029 \\
\citet{Gordon:2023ck}  & -0.368 $\pm$ 0.248 & 0.818 $\pm$ 0.457 & -0.100 $\pm$ 0.051 & -0.043 $\pm$ 0.080 \\
\enddata
\tablenotetext{a}{This is what is adopted throughout the paper.}
\tablecomments{Description of columns.}
\end{deluxetable*}

\section{HST-Gaia Proper Motion Catalog in SgrA*-at-Rest Coordinates}
\label{app:sgra_pm}
For convenience, the catalog of HST-Gaia proper motion measurements that have been transformed into the SgrA*-at-Rest coordinate system (as discussed in $\mathsection$\ref{sec:sgra_rest}) is presented in Table \ref{tab:sgra_pm}.

\movetableright=3mm
\begin{deluxetable*}{lccccccccccccc}
\tablewidth{0pt}
\tabletypesize{\tiny}
\tablecaption{HST-Gaia Proper Motions: SgrA*-at-Rest Coordinates}
\tablehead{
\colhead{Name} & \colhead{$\delta_{\alpha^*_0}^{s}$} & \colhead{$\sigma_{\delta_{\alpha^*_0}^{s}}$} & \colhead{$\delta_{\delta_0}^{s}$} & \colhead{$\sigma_{\delta_{\delta_0}^{s}}$} &
\colhead{$\mu_{\alpha*}^{s}$} & \colhead{$\sigma_{\mu_{\alpha*}^{s}}$} & \colhead{$\mu_{\delta}^{s}$} & \colhead{$\sigma_{\mu_{\delta}^{s}}$} & \colhead{$t_0^{s}$} &
\colhead{Ref Type} & \colhead{Alt Name}\\
& ($''$) & (mas) & ($''$) & (mas) &  (mas/yr) & (mas/yr) & (mas/yr) & (mas/yr) & (year) &
& &
}
\startdata
HST\_NSC\_007314 & 0.0137 & 3.76 & -0.2925 & 3.84 & -5.81 & 0.90 & 10.05 & 0.89 & 2018.4467 & 0 & S0-6 \\
HST\_NSC\_007079 & 0.2626 & 1.01 & -0.6155 & 1.00 & 8.02 & 0.25 & -5.15 & 0.25 & 2015.4906 & 0 & S0-9 \\
HST\_NSC\_006757 & 0.5422 & 0.43 & -0.3986 & 0.42 & 3.31 & 0.18 & -0.85 & 0.18 & 2019.6907 & 0 & S0-13 \\
HST\_NSC\_007316 & -0.5360 & 5.06 & 0.4391 & 1.06 & 0.90 & 0.94 & 2.47 & 0.30 & 2019.3942 & 0 & S0-12 \\
HST\_NSC\_007214 & -0.7304 & 0.40 & -0.3036 & 0.46 & 2.95 & 0.08 & -1.90 & 0.10 & 2018.5698 & 0 & S0-14 \\
HST\_NSC\_007322 & 0.3089 & 1.25 & 0.8443 & 1.25 & -8.43 & 0.38 & 3.00 & 0.38 & 2012.8189 & 0 & S1-3 \\
HST\_NSC\_006751 & 0.3076 & 5.72 & -0.8739 & 5.27 & -3.40 & 0.92 & 4.50 & 0.88 & 2013.5218 & 0 & S1-5 \\
HST\_NSC\_016387 & -0.8275 & 9.01 & 0.4384 & 2.66 & 7.57 & 1.83 & -1.68 & 0.66 & 2017.8891 & 0 & S1-26 \\
HST\_NSC\_007323 & 1.0858 & 0.84 & 0.0458 & 0.58 & 5.16 & 0.16 & 0.94 & 0.13 & 2017.6056 & 0 & S1-1 \\
HST\_NSC\_006701 & 0.9668 & 0.24 & 0.6109 & 0.22 & -9.44 & 0.05 & 7.04 & 0.04 & 2018.6155 & 0 & irs16C \\
\enddata
\tablecomments{Description of columns: \emph{Name}: star name,
\emph{$\delta_{\alpha^*_0}^{s}$, $\delta_{\delta_0}^{s}$}: star position at $t_0^{s}$,
\emph{$\sigma_{\delta_{\alpha^*_0}^{s}}$, $\sigma_{\delta_{\delta_0}^{s}}$}: star position error at $t_0^{s}$,
\emph{$\mu_{\alpha*}^{sgra}$, $\mu_{\delta}^s$}: SgrA*-at-rest proper motions,
\emph{$\sigma_{\mu_{\alpha*}^{s}}$, $\sigma_{\mu_{\delta}^{s}}$}: error in SgrA*-at-Rest proper motions,
\emph{$t_0^{s}$}: reference epoch of the SgrA*-at-rest proper motions (calculated as the astrometric-error weighted average time across the epoch dates in Table \ref{tab:obs}),
\emph{Ref Type}: Type of ref star -- 0: Secondary reference star, 1: primary reference star, 2: maser,
\emph{Alt Name}: Alternative name for star, if applicable \citep[either from][or Gaia-DR3 catalog]{Sakai:2019fm}
}
\tablecomments{The proper motion fit uncertainties reported in this table ($\sigma_{\delta_{\alpha^*_0}^{s}}$, $\sigma_{\delta_{\delta_0}^{s}}$, $\sigma_{\mu_{\delta}}$, $\sigma_{\mu_{\alpha*}^{s}}$, $\sigma_{\mu_{\delta}^{s}}$)
do not include the systematic error between the HST-Gaia and the SgrA*-at-Rest frame discussed in $\mathsection$\ref{sec:icrf3_sgra}.}
\tablecomments{This table is available in its entirety in the machine-readable format.}
\label{tab:sgra_pm}
\end{deluxetable*}

\end{document}